\documentclass[twocolumn,tighten,times]{aastex63} %,twocolappendix ,linenumbers

\usepackage{amsmath} %%-->for use of eqref
\usepackage{amssymb}
\accepted{May 15, 2022}
\newcommand{\galprop}{\textsc{GalProp}}
\newcommand{\helmod}{\textsc{HelMod}}

\newcommand{\gray}{$\gamma$-ray}

\shorttitle{
Spectra of Cosmic Ray Sodium and Aluminum}
\shortauthors{Boschini et al.}

\begin{document}

\title{
Spectra of Cosmic Ray Sodium and Aluminum and Unexpected Aluminum Excess
}

\author[0000-0002-6401-0457]{M.~J.~Boschini}
\affiliation{INFN, Milano-Bicocca, Milano, Italy}
\affiliation{CINECA, Segrate, Milano, Italy}

\author[0000-0002-7669-0859]{S.~{Della~Torre}}
\affiliation{INFN, Milano-Bicocca, Milano, Italy}

\author[0000-0003-3884-0905]{M.~Gervasi}
\affiliation{INFN, Milano-Bicocca, Milano, Italy}
\affiliation{Physics Department, University of Milano-Bicocca, Milano, Italy}

\author[0000-0003-1942-8587]{D.~Grandi}
\affiliation{INFN, Milano-Bicocca, Milano, Italy}
\affiliation{Physics Department, University of Milano-Bicocca, Milano, Italy}

\author[0000-0003-1458-7036]{G.~J\'{o}hannesson} 
\affiliation{Science Institute, University of Iceland, Dunhaga 3, IS-107 Reykjavik, Iceland}
\affiliation{NORDITA,  Roslagstullsbacken 23, 106 91 Stockholm, Sweden}

\author[0000-0002-2168-9447]{G.~{La~Vacca}}
\affiliation{INFN, Milano-Bicocca, Milano, Italy}
\affiliation{Physics Department, University of Milano-Bicocca, Milano, Italy}

\author[0000-0002-3729-7608]{N.~Masi}
\affiliation{INFN, Bologna, Italy}
\affiliation{Physics Department, University of Bologna, Bologna, Italy}

\author[0000-0001-6141-458X]{I.~V.~Moskalenko} 
\affiliation{Hansen Experimental Physics Laboratory, Stanford University, Stanford, CA 94305}
\affiliation{Kavli Institute for Particle Astrophysics and Cosmology, Stanford University, Stanford, CA 94305}

\author{S.~Pensotti}
\affiliation{INFN, Milano-Bicocca, Milano, Italy}
\affiliation{Physics Department, University of Milano-Bicocca, Milano, Italy}

\author[0000-0002-2621-4440]{T.~A.~Porter} 
\affiliation{Hansen Experimental Physics Laboratory, Stanford University, Stanford, CA 94305}
\affiliation{Kavli Institute for Particle Astrophysics and Cosmology, Stanford University, Stanford, CA 94305}

\author{L.~Quadrani}
\affiliation{INFN, Bologna, Italy}
\affiliation{Physics Department, University of Bologna, Bologna, Italy}

\author[0000-0002-1990-4283]{P.~G.~Rancoita}
\affiliation{INFN, Milano-Bicocca, Milano, Italy}

\author[0000-0002-7378-6353]{D.~Rozza}
\affiliation{INFN, Milano-Bicocca, Milano, Italy}
%\affiliation{Physics Department, University of Milano-Bicocca, Milano, Italy}

\author[0000-0002-9344-6305]{M.~Tacconi}
\affiliation{INFN, Milano-Bicocca, Milano, Italy}
\affiliation{Physics Department, University of Milano-Bicocca, Milano, Italy}
%\affil{institute}
%\email{\myemail}



%$_{\phantom{1}9}^{19}$F 

\begin{abstract}

Since its launch, the Alpha Magnetic Spectrometer--02 (AMS-02) has delivered outstanding quality measurements of the spectra of cosmic-ray (CR) species, $\bar{p}$, $e^{\pm}$, and nuclei (H--Si, Fe), which resulted in a number of breakthroughs. The most recent AMS-02 result is the measurement of the spectra of CR sodium and aluminum up to $\sim$2 TV. Given their low solar system abundances, a significant fraction of each element is produced in fragmentations of heavier species, predominantly Ne, Mg, and Si. In this paper, we use precise measurements of the sodium and aluminum spectra by AMS-02 together with ACE-CRIS and Voyager 1 data to test their origin. We show that the sodium spectrum agrees well with the predictions made with the \galprop{}-\helmod{} framework, while aluminum spectrum shows a significant excess in the rigidity range from 2--7 GV. In this context, we discuss the origin of other low-energy excesses in Li, F, and Fe found earlier. The observed excesses in Li, F, and Al appear to be consistent with the local Wolf-Rayet (WR) stars hypothesis, invoked to reproduce anomalous $^{22}$Ne/$^{20}$Ne, $^{12}$C/$^{16}$O, and $^{58}$Fe/$^{56}$Fe ratios in CRs, while excess in Fe is likely connected with a past SN activity in the solar neighborhood. We also provide updated local interstellar spectra (LIS) of sodium and aluminum in the rigidity range from few MV to $\sim$2 TV. Our calculations employ the self-consistent \galprop{}--\helmod{} framework that has proved to be a reliable tool in deriving the LIS of CR $\bar{p}$, $e^{-}$, and nuclei $Z\le28$.
\end{abstract}



\section{Introduction} \label{Intro}
%%%%%%%%%%%%%%%%%%%%%%%%%%%%%%%%%%%%%%%%%%%%%%%%%%%
%%%%%%%%%%%%%%%%%%%%%%%%%%%%%%%%%%%%%%%%%%%%%%%%%%%

Precise data at low and high energies delivered by the modern space instrumentation have triggered a number of discoveries of new features in spectra of CR species enabling unprecedented probes of the stellar nucleosynthesis, properties of the interstellar medium (ISM), and the origin of CRs. A combination of data from individual spacecraft, can be combined to cover the enormous range of rigidities, from few MV to tens of TV, where the individual spectra of CR species are shaped by many different processes. 

With their latest paper on CR sodium and aluminum, the AMS-02 collaboration \citep{2021PhRvL.126h1102A} has now completed their series of papers on CR species H--Si, and Fe \citep{2014PhRvL.113v1102A, 2015PhRvL.114q1103A, 2015PhRvL.115u1101A, 2016PhRvL.117i1103A, 2016PhRvL.117w1102A, 2017PhRvL.119y1101A, 2018PhRvL.120b1101A, 2018PhRvL.121e1103A, 2019PhRvL.122d1102A, 2019PhRvL.122j1101A, 2020PhRvL.124u1102A, 2021PhRvL.126d1104A, 2021PhRvL.126h1102A, 2021PhRvL.127b1101A}. Fluorine, sodium, and aluminum in CRs share some common features: all three have odd atomic numbers, and are mostly represented by a single stable isotope with odd number of nucleons, $^{19}_{\phn9}$F, $^{23}_{11}$Na, $^{27}_{13}$Al, while long-lived radioactive $^{26}$Al is present in CRs only fractionally. Because of the properties of stellar nucleosynthesis, the source and CR abundances of these elements are considerably lower than their even-$Z$ neighbors, $_{8}$O, $_{10}$Ne, $_{12}$Mg, $_{14}$Si. A significant fraction of each element, F, Na, Al, is produced through fragmentation of their more abundant neighbors, Ne, Mg, Si. A comparison of the spectra of F, Na, and Al, their common and distinct features, may provide new insights into the origin of these elements.

In this paper we analyze the new CR measurements of the spectra of Na and Al and test their consistency with measurements of other species. We also provide updated Na and Al LIS in the rigidity range from few MV to $\sim$2 TV. Our calculations and interpretation employ the \galprop{}\footnote{Available from http://galprop.stanford.edu \label{galprop-site}}--\helmod{}\footnote{http://www.helmod.org/ \label{helmod-footnote}} framework that is proved to be a reliable tool in deriving the LIS of CR species \citep{2019HelMod, 2020ApJS..250...27B}.

\section{Calculations} \label{calcs}
%%%%%%%%%%%%%%%%%%%%%%%%%%%%%%%%%%%%%%%%%%%%%%%%%%%
%%%%%%%%%%%%%%%%%%%%%%%%%%%%%%%%%%%%%%%%%%%%%%%%%%%

In this work we are using the same CR propagation model with distributed reacceleration and convection that was used in our previous analyses \citep[for more details see][]{2017ApJ...840..115B,  2018ApJ...854...94B, 2018ApJ...858...61B, 2020ApJS..250...27B, 2020ApJ...889..167B, 2021ApJ...913....5B, 2022ApJ...925..108B}. The latest version 57 of the \galprop{} code for Galactic propagation of CRs and diffuse emissions is described in detail in \citet{2021arXiv211212745P}, see also \citet{2020ApJS..250...27B} and references therein.

Full details of the latest \helmod{} code version 4 for heliospheric propagation are provided in \citet{2019AdSpR..64.2459B}. It solves the Fokker-Planck equation for heliospheric propagation in Kolmogorov formulation backward in time \citep{2016JGRA..121.3920B}. The accuracy of the solution was tested using the Crank-Nicholson technique and found to be better than 0.5\% at low rigidities. The large number of simulated events ensures that the statistical errors are negligible compared to the other modeling uncertainties.

When comparing our calculations with data collected over extended period of time, variations in the solar activity are addressed in the following way. The propagation equation is solved for each Carrington rotation, and the numerical results are then combined accordingly to the instrument exposure and the time period. This approach is equivalent to application of a weighted average that accounts for both exposure time and absolute counting rate variations.

The values of propagation parameters in the ISM along with their confidence limits are derived from the best available CR data using the Markov Chain Monte Carlo (MCMC) routine. Here we use the same method as described in \citet{2017ApJ...840..115B}. Five main propagation parameters, that affect the overall shape of CR spectra, were left free in the scan using \galprop{} running in 2D mode: the Galactic halo half-width $z_h$, the normalization of the diffusion coefficient $D_0$ at the reference rigidity $R=4$ GV and the index of its rigidity dependence $\delta$, the Alfv\'en velocity $V_{\rm Alf}$, and the gradient of the convection velocity $dV_{\rm conv}/dz$ ($V_{\rm conv}=0$ in the plane, $z=0$). Their best-fit values tuned to the AMS-02 data are listed in Table~\ref{tbl-prop} and are the same as obtained in \citet{2020ApJS..250...27B}. The radial size of the Galaxy does not significantly affect the values of propagation parameters and was set to 20 kpc. We also introduced a factor $\beta^\eta$ in the diffusion coefficient, where $\beta=v/c$, and $\eta$ was left free. The best fit value of $\eta=0.70$ improves the agreement at low energies, and slightly affects the choice of injection indices ${\gamma}_0$ an ${\gamma}_1$. A detailed discussion of the injection ({\it I}) and propagation ({\it P}) scenarios of the 350 GV break can be found in the works by \citet{2012ApJ...752...68V} and \citet{2020ApJS..250...27B}.

The corresponding B/C ratio also remains the same \citep[see Fig.~4 of][]{2020ApJS..250...27B}, and compares well with all available measurements: Voyager~1 \citep{2016ApJ...831...18C}, ACE-CRIS\footnote{http://www.srl.caltech.edu/ACE/ASC/level2/cris\_l2desc.html} \citep{2013ApJ...770..117L}, AMS-02 \citep{2018PhRvL.120b1101A}, ATIC-2 \citep{2009BRASP..73..564P}, CREAM \citep{2008APh....30..133A, 2009ApJ...707..593A}, and NUCLEON \citep{2019AdSpR..64.2559G}. 

In the calculations, we use our standard formalism. The convection velocity is assumed to increase linearly with distance $z$ from the plane, $V_{\rm conv}(z)=V_0+ z\cdot dV_{\rm conv}/dz$. The spatial diffusion coefficient is parametrized as $D_{xx} = \beta D_0 R^\delta$, where $\beta=v/c$ is the particle velocity. If reacceleration is included, the momentum-space diffusion coefficient $D_{pp}$ is related to $D_{xx}$ as $D_{pp}=p^{2}V_{\rm Alf}^{2}/\left(  9D_{xx}\right)$ \citep{1990acr..book.....B, 1994ApJ...431..705S}. See also a review by \citet{2007ARNPS..57..285S} and our papers cited in the beginning of this section. The injection spectra of CR species are parametrized by the rigidity-dependent function:
\begin{equation}  \label{eq:1}
q(R) \propto (R/R_0)^{-\gamma_0}\prod_{i=0}^2\bigg[1 + (R/R_i)^\frac{\gamma_i - \gamma_{i+1}}{s_i}\bigg]^{s_i},
\end{equation}
where $\gamma_{i =0,1,2,3}$ are the spectral indices, $R_{i = 0,1,2}$ are the break rigidities, $s_i$ are the smoothing parameters ($s_i$ is negative/positive for $|\gamma_i |\lessgtr |\gamma_{i+1} |$). 

\begin{deluxetable}{rlcc}[tb!]
	\def\arraystretch{0.9}
	\tablewidth{0mm}
	\tablecaption{Best-fit propagation parameters for {\it I}- and {\it P}-scenarios\label{tbl-prop}}
	\tablehead{
		\colhead{Parameter}& \multicolumn{1}{l}{Units}& \colhead{Best Value}& \colhead{Error} 
	}
	\startdata
	$z_h$ & kpc &4.0 &0.6\\
	$D_0 (R= 4\ {\rm GV})$ & cm$^{2}$ s$^{-1}$  & $4.3\times10^{28}$ &0.7\\
	$\delta$\tablenotemark{a} & &0.415 &0.025\\
	$V_{\rm Alf}$ & km s$^{-1}$ &30 &3\\
	$dV_{\rm conv}/dz$ & km s$^{-1}$ kpc$^{-1}$ & 9.8 &0.8
	\enddata
	\tablenotetext{a}{The {\it P}-scenario assumes a break in the diffusion coefficient with index $\delta_1=\delta$ below the break and index $\delta_2=0.15\pm 0.03$ above the break at $R=370\pm 25$ GV \citep[for details see][]{2020ApJ...889..167B}.}
\end{deluxetable}

\section{Results and Discussion} \label{results}
%%%%%%%%%%%%%%%%%%%%%%%%%%%%%%%%%%%%%%%%%%%%%%%%%%%
%%%%%%%%%%%%%%%%%%%%%%%%%%%%%%%%%%%%%%%%%%%%%%%%%%%

\begin{figure*}[tb!]
	\centering
	\includegraphics[width=0.49\textwidth]{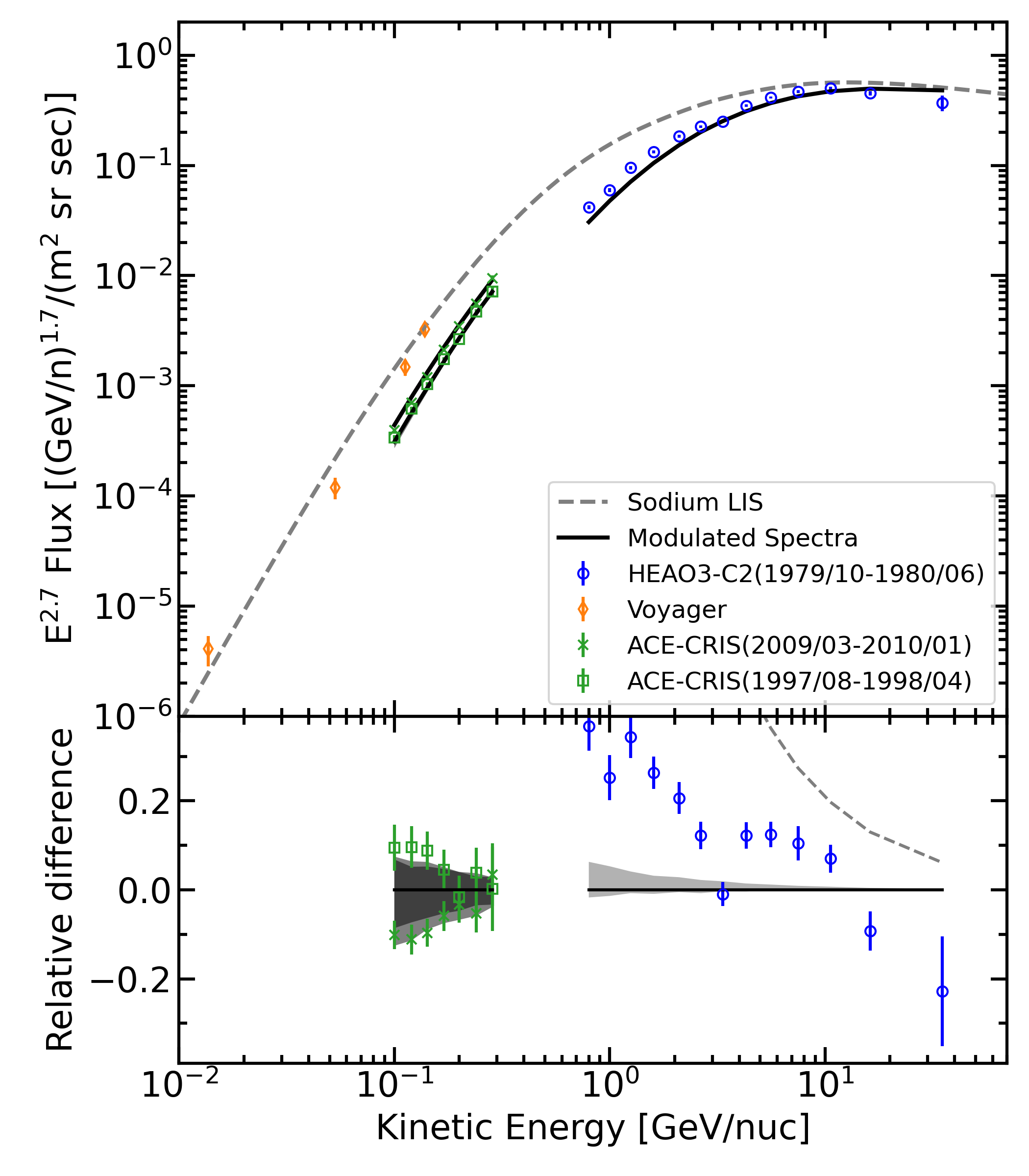}
	\includegraphics[width=0.49\textwidth]{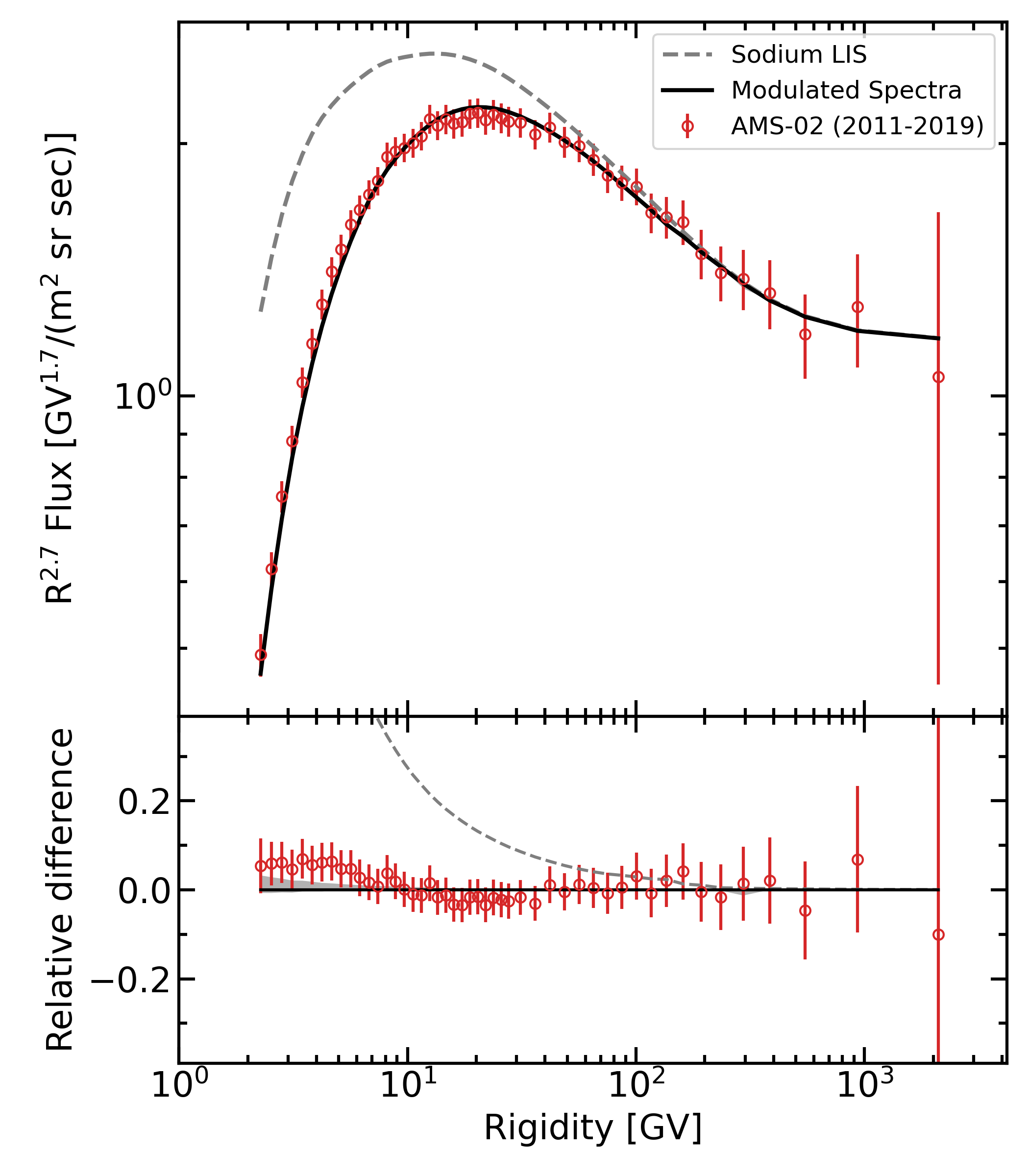}
	\caption{	
A comparison of the calculated spectrum of sodium with Voyager~1 \citep{2016ApJ...831...18C}, ACE-CRIS \citep{2013ApJ...770..117L}, HEAO-3-C2 \citep{1990A&A...233...96E} (left panel), and AMS-02 data \citep{2021PhRvL.127b1101A} (right panel). In the left panel, to match the units of the published data, the spectra are plotted vs.\ kinetic energy per nucleon, $E_{\rm kin}$. The dashed gray lines show the LIS, and the solid black lines are the corresponding modulated spectra. The lower panels show the relative difference between our calculations and the data sets. The gray shaded areas indicate 1$\sigma$ confidence limits for the calculated modulated spectra.
	}
	\label{fig:Na-spec}
\end{figure*}

Fig.~\ref{fig:Na-spec} shows a comparison of the calculated spectrum with Voyager~1 \citep{2016ApJ...831...18C}, ACE-CRIS \citep{2013ApJ...770..117L}, HEAO-3-C2 \citep{1990A&A...233...96E}, and AMS-02 sodium data \citep{2021PhRvL.127b1101A}. The calculated spectrum reproduces the data in the whole energy range from 100 MV--2 TV quite well, although there is insignificant 1$\sigma$ excess below 7 GV. The injection spectrum of primary sodium is adjusted to the AMS-02 data (Table~\ref{tbl-inject}), but remains similar to other species \citep{2020ApJS..250...27B}. 

\begin{deluxetable*}{ccc rcl rcl rcl r}[tb!]        
	\tabletypesize{\footnotesize}
        \def\arraystretch{1.1}
        \tablecolumns{13}
        \tablewidth{0mm}
        \tablecaption{The injection spectra of primary sodium and aluminum and aluminum excess \label{tbl-inject}}
        \tablehead{
                \multicolumn{2}{c}{} & 
                \multicolumn{1}{c}{Source} &
                \multicolumn{10}{c}{Spectral parameters}\\
                \cline{4-13}
                \multicolumn{2}{c}{} & 
                \multicolumn{1}{c}{abundance} & 
                \multicolumn{1}{c}{$\gamma_0$} & \multicolumn{1}{c}{$R_0$ (GV)} & \multicolumn{1}{l}{$s_0$} &
                \multicolumn{1}{c}{$\gamma_1$} & \multicolumn{1}{c}{$R_1$ (GV)} & \multicolumn{1}{l}{$s_1$} &  
                \multicolumn{1}{c}{$\gamma_2$} & \multicolumn{1}{c}{$R_2$ (GV)} & \multicolumn{1}{l}{$s_2$} &  
                \multicolumn{1}{c}{$\gamma_3$} 
                 }
	\startdata
$^{23}$Na & default & 50&0.24 & 0.92 & 0.22 & 2.33 & 8.2 & 0.20 & 2.49 & 350 & 0.16 & 2.15\\
$^{27}$Al & default & 94&0.20 & 0.60 & 0.17 & 2.04 & 7.0 & 0.20 & 2.46 & 355 & 0.17 & 1.93\\
$^{27}$Al & excess & 11.5&$-5.00$ & 4.0 & 0.30 & 5.00 & \nodata & \nodata & \nodata & \nodata & \nodata & \nodata \\
	\enddata
  \tablecomments{The source abundances are relative, see Table 3 in \citet{2020ApJS..250...27B}. For parameter definitions see Eq.~(\ref{eq:1}). Shown are $|s_i|$ values, note that $s_i$ is negative/positive for $|\gamma_i |\lessgtr |\gamma_{i+1} |$.}
\end{deluxetable*}

\begin{figure*}[tbh!]
	\centering
	\includegraphics[width=0.49\textwidth]{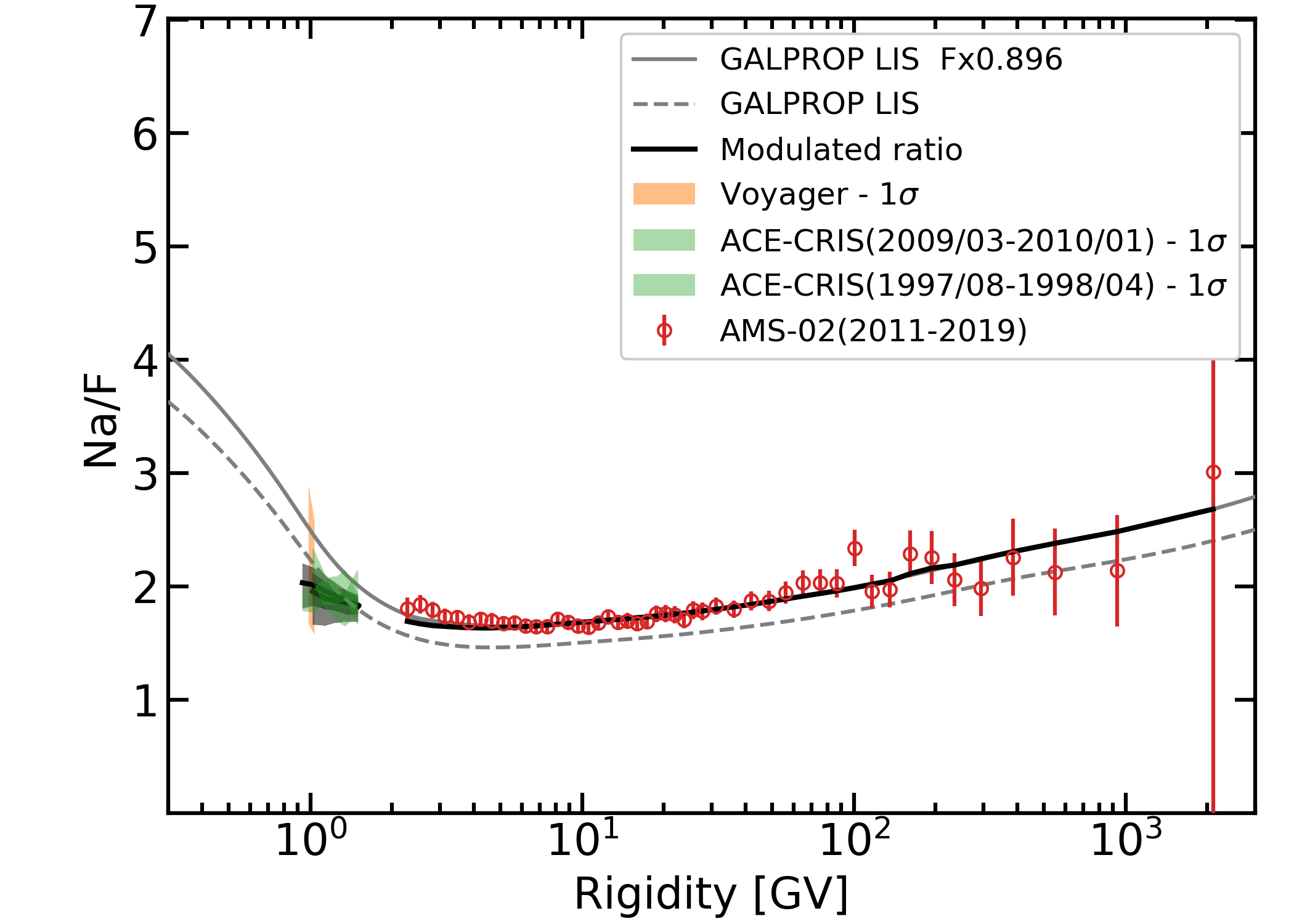}\hfill
	\includegraphics[width=0.49\textwidth]{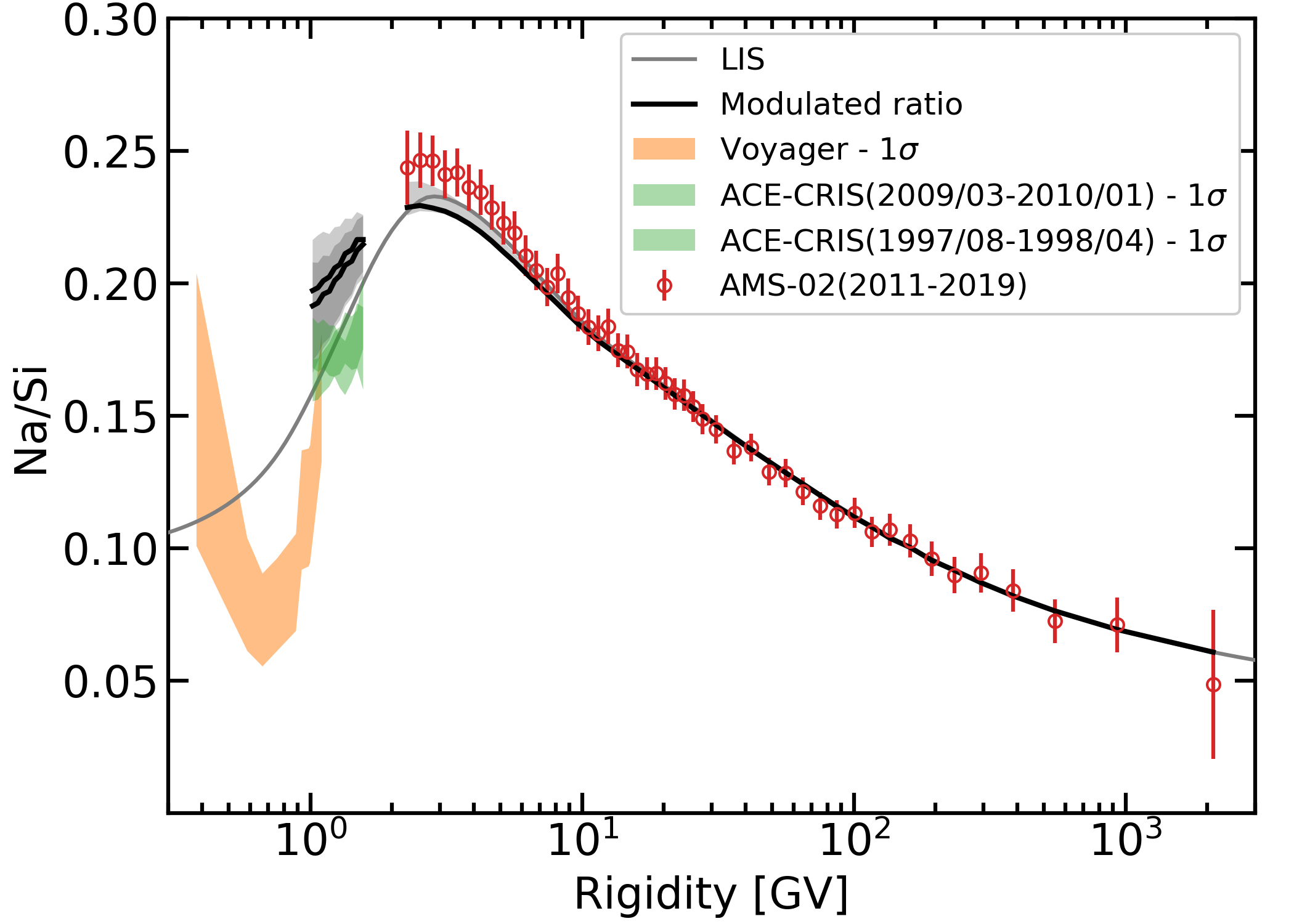}
	\caption{
The calculated Na/F and Na/Si ratios as compared with Voyager~1 \citep{2016ApJ...831...18C}, ACE-CRIS \citep{2013ApJ...770..117L}, and AMS-02 data \citep{2020PhRvL.124u1102A, 2021PhRvL.126h1102A, 2021PhRvL.127b1101A}.
In the left panel, the dashed gray line shows the default LIS Na/F ratio, while the solid gray line shows the ratio for the renormalized ($\times$0.896) fluorine spectrum \citep[for details see][]{2022ApJ...925..108B}. The solid gray line in the right panel shows the LIS Na/Si ratio. In both panels, the modulated ratios are shown with solid black lines. The gray shaded areas indicate 1$\sigma$ confidence limits for the calculated modulated ratios.
%corresponding to the ACE-CRIS data taken during consequent solar minima. 
The Voyager 1 and ACE-CRIS data are converted from kinetic energy per nucleon to rigidity (assuming $A/Z=2$ for Si). These data are shown as shaded areas with the width corresponding to 1$\sigma$ error.
	}
	\label{fig:Na-F-Si-ratio}
\end{figure*}

\begin{figure*}[tbh!]
	\centering
	\includegraphics[width=0.49\textwidth]{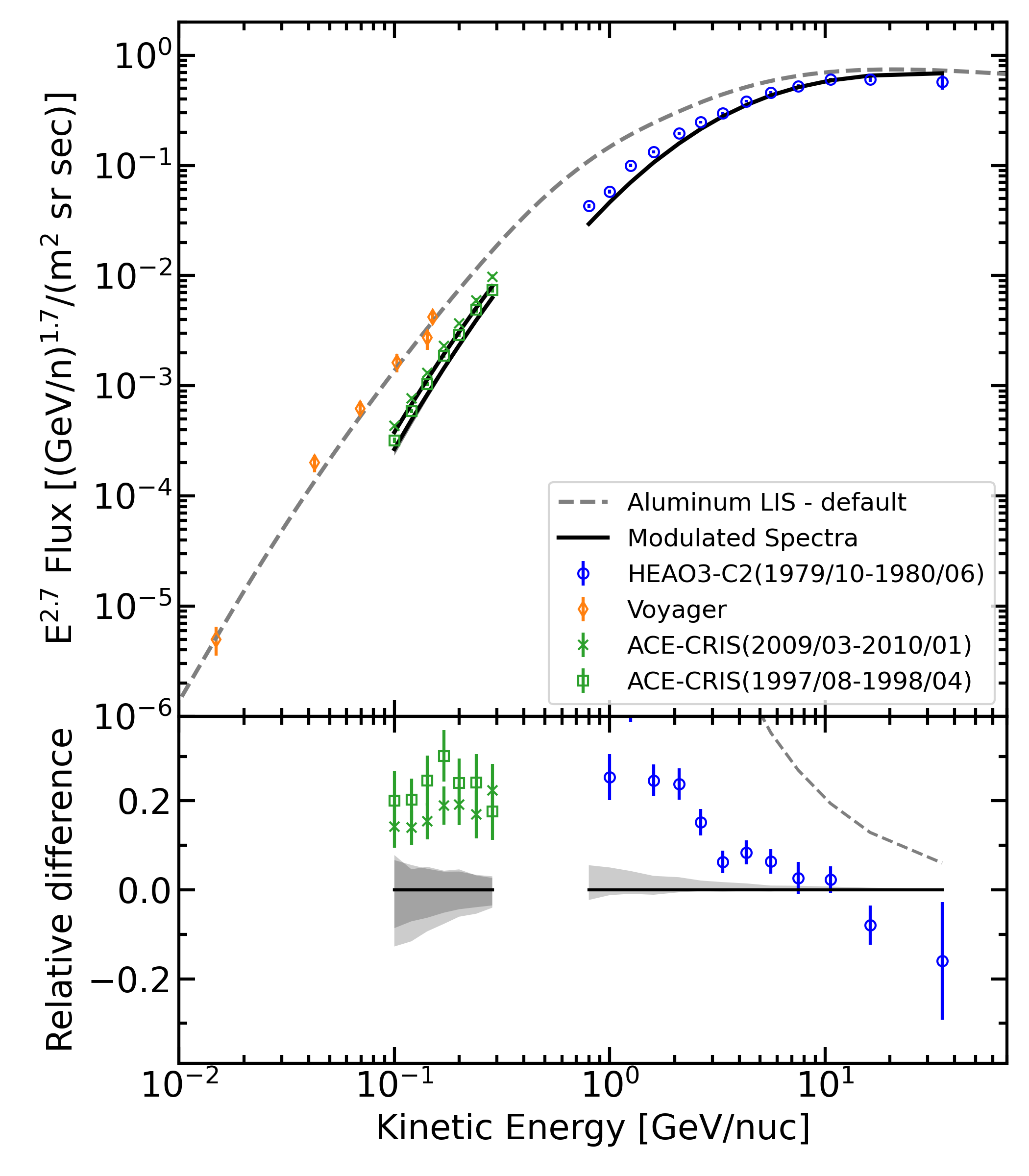}\hfill
	\includegraphics[width=0.49\textwidth]{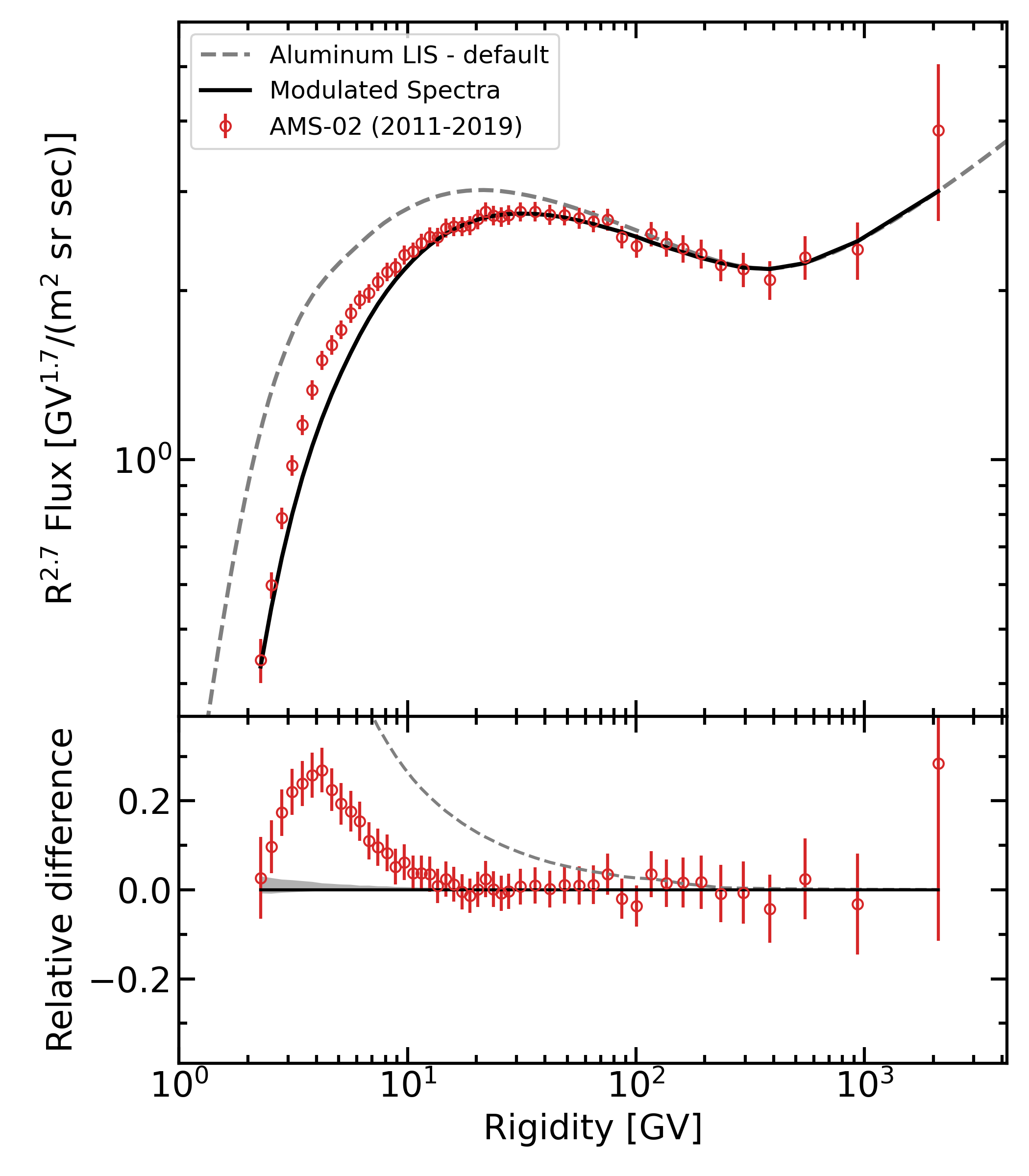}\\
	\includegraphics[width=0.49\textwidth]{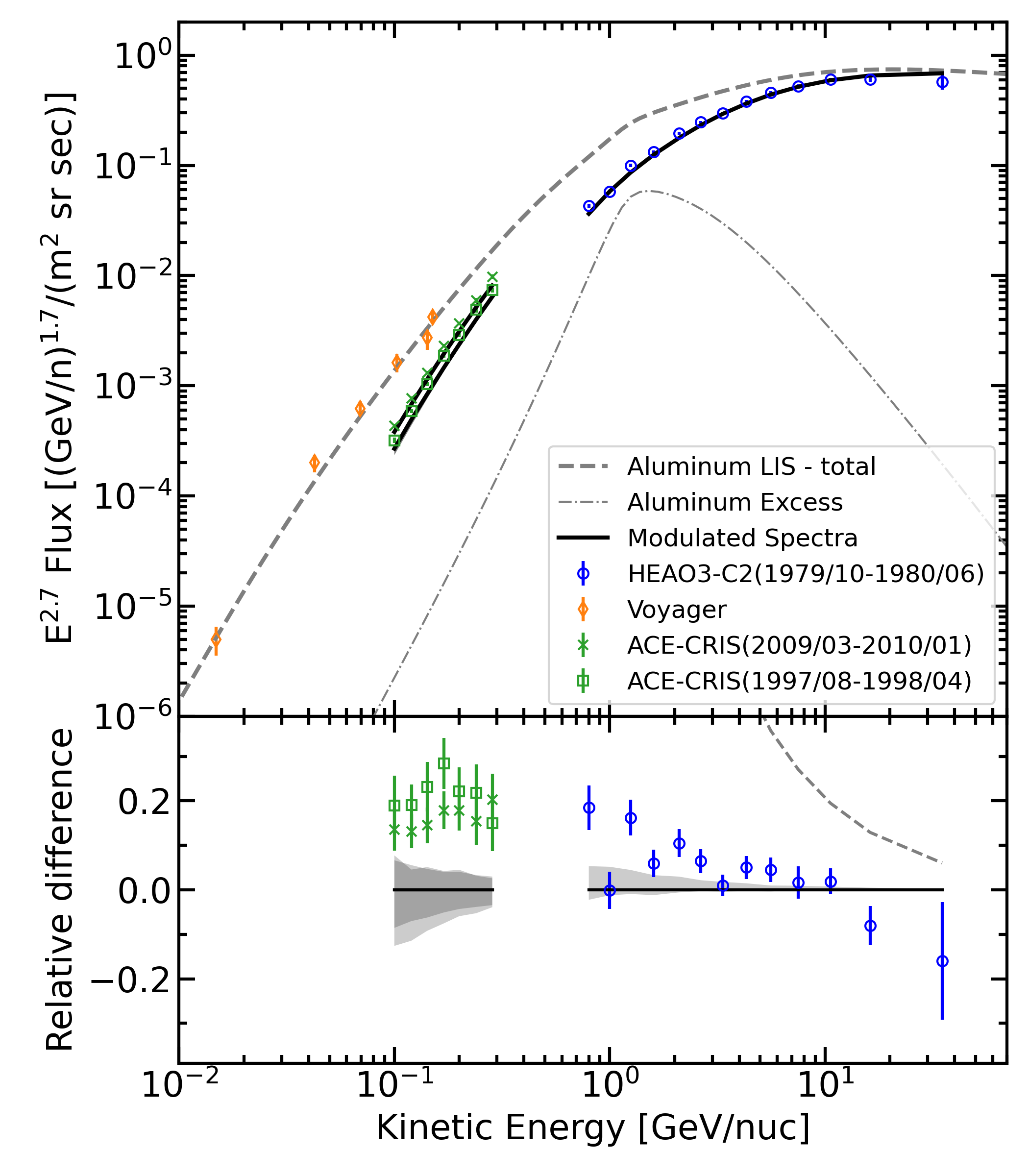}\hfill
	\includegraphics[width=0.49\textwidth]{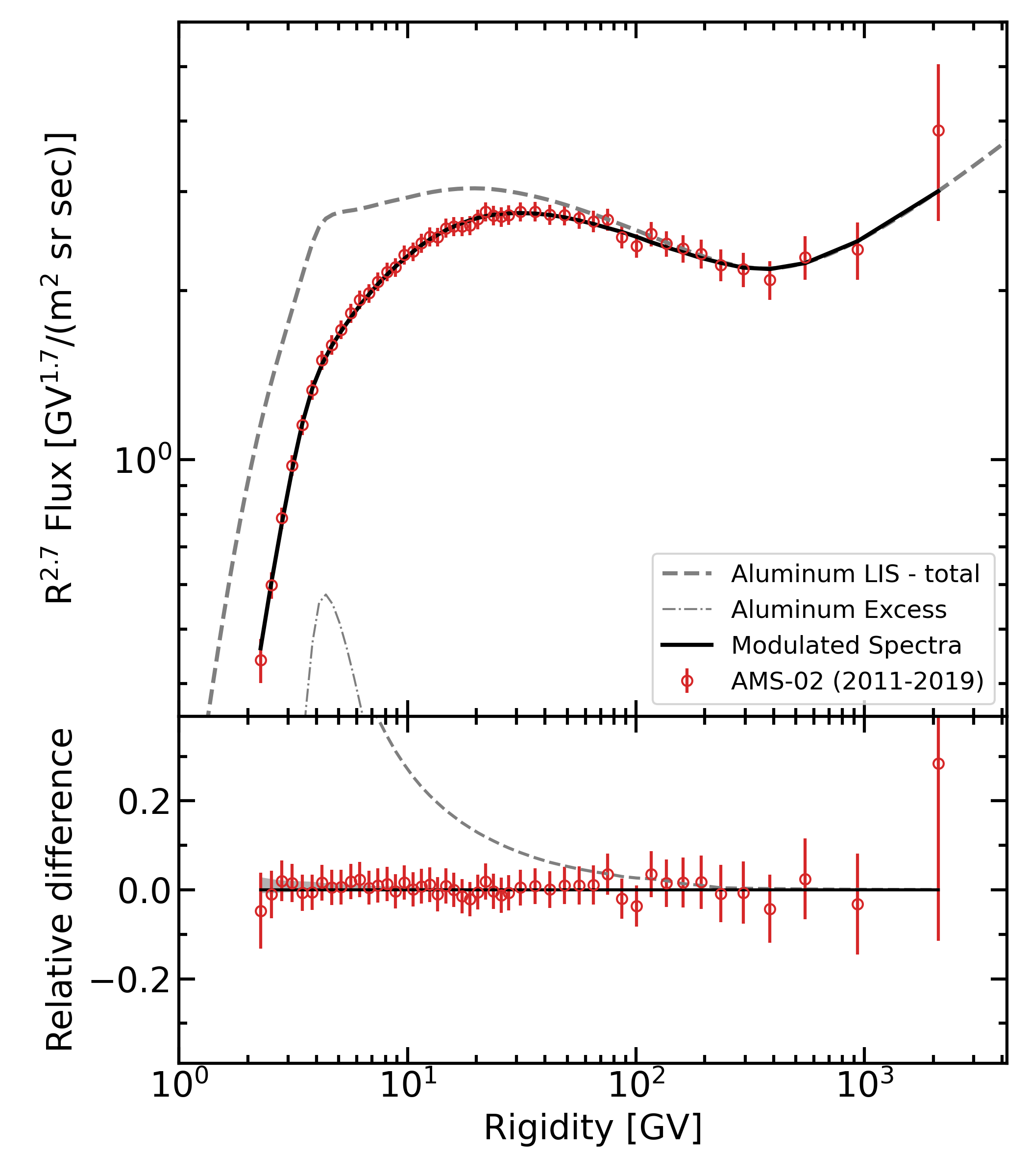}
	\caption{
A comparison of the calculated aluminum spectrum with Voyager~1 \citep{2016ApJ...831...18C}, ACE-CRIS \citep{2013ApJ...770..117L}, HEAO-3-C2 \citep{1990A&A...233...96E}, and AMS-02 data \citep{2021PhRvL.127b1101A}. 
Top panels show the calculations made with the default aluminum injection spectrum (Table \ref{tbl-inject}) similar to other Si-group species \citep[Table 2,][]{2020ApJS..250...27B}. The bottom panels show the total LIS (default+excess), with the injection spectrum of the aluminum excess (Table \ref{tbl-inject}) tuned to the AMS-02 data. The excess is shown separately with the dash-dotted gray lines. In the left panels, to match the units of the published data, the spectra are plotted vs.\ kinetic energy per nucleon, $E_{\rm kin}$.
In all panels, the dashed gray line shows the LIS, and the solid black line is the corresponding modulated spectrum. The gray shaded areas indicate 1$\sigma$ confidence limits for the calculated modulated spectrum. 
The lower panels show the relative difference between our calculations and the data sets. 
	} \vspace{4\baselineskip}
	\label{fig:Al-spec}
\end{figure*}

\begin{figure*}[!tb]
	\centering
	\includegraphics[width=0.49\textwidth]{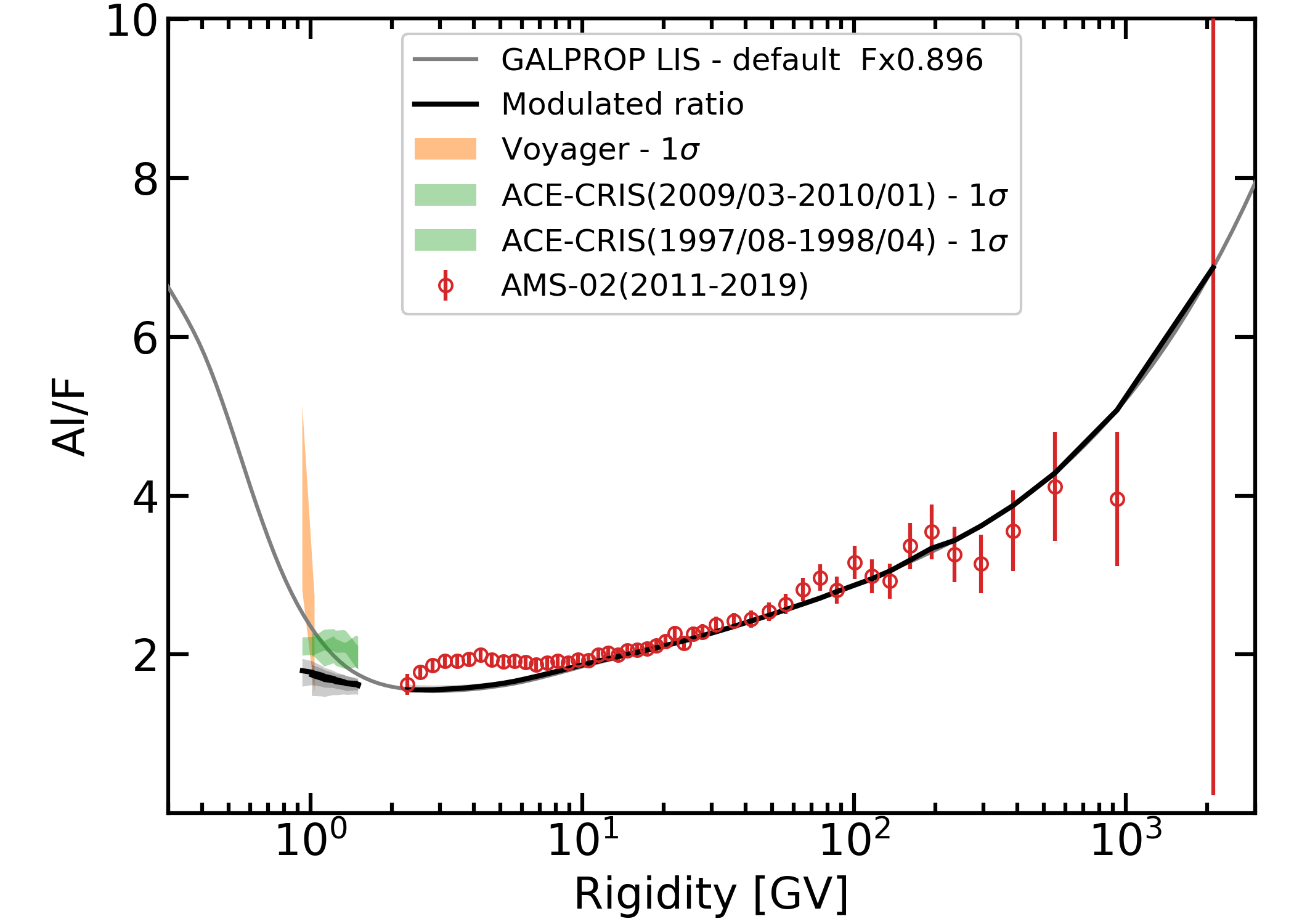}\hfill
	\includegraphics[width=0.49\textwidth]{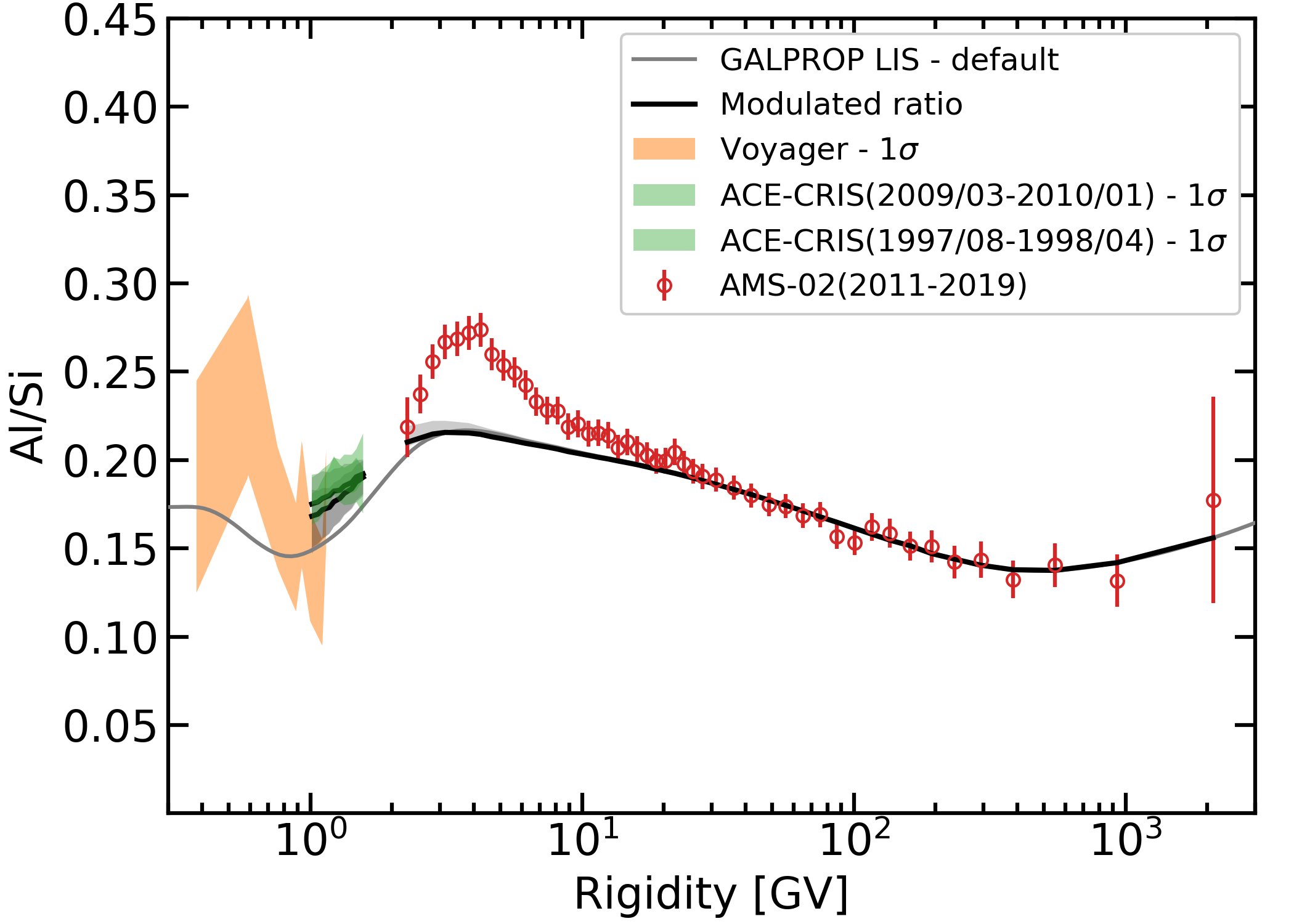}\\
	\includegraphics[width=0.49\textwidth]{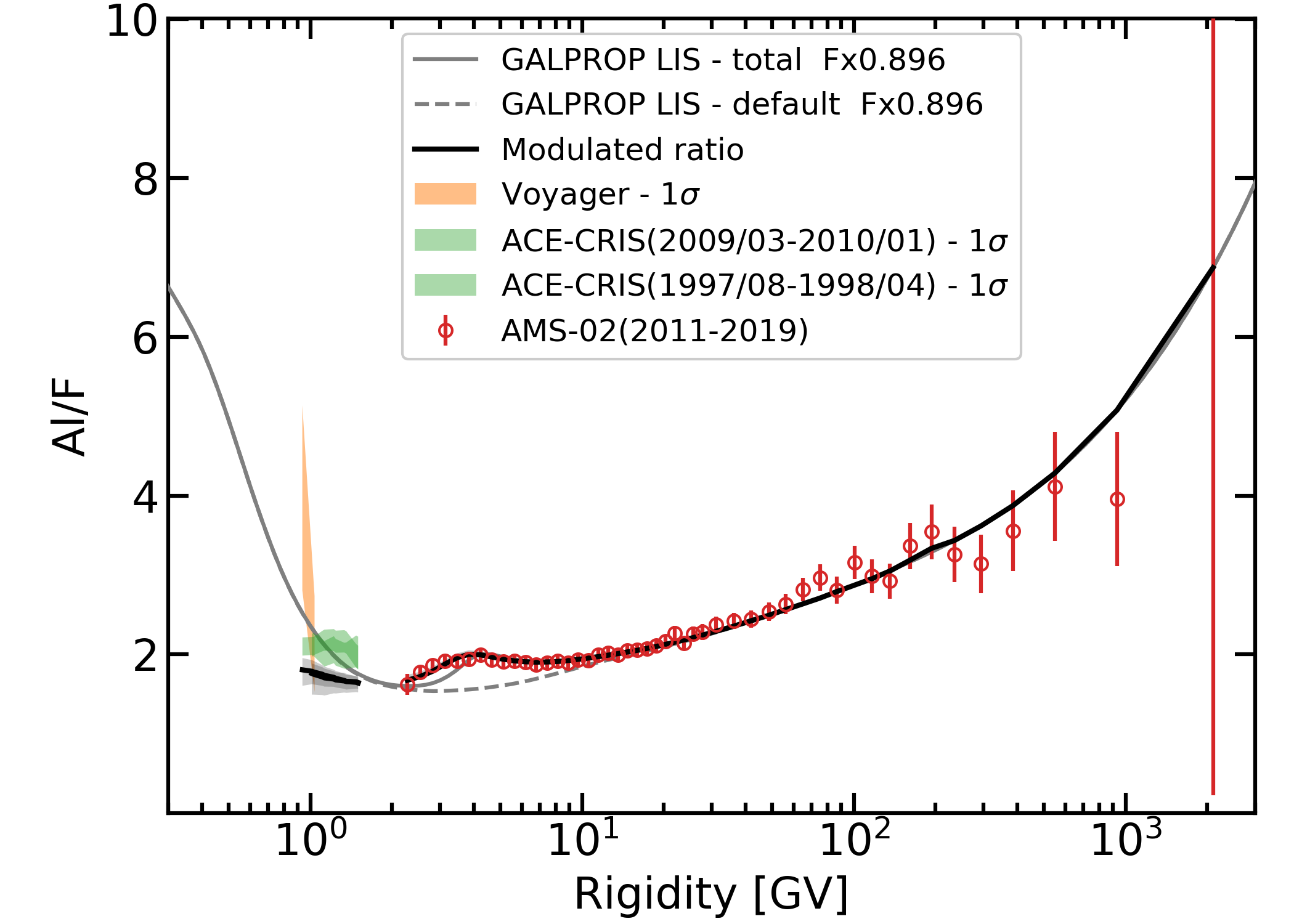}\hfill
	\includegraphics[width=0.49\textwidth]{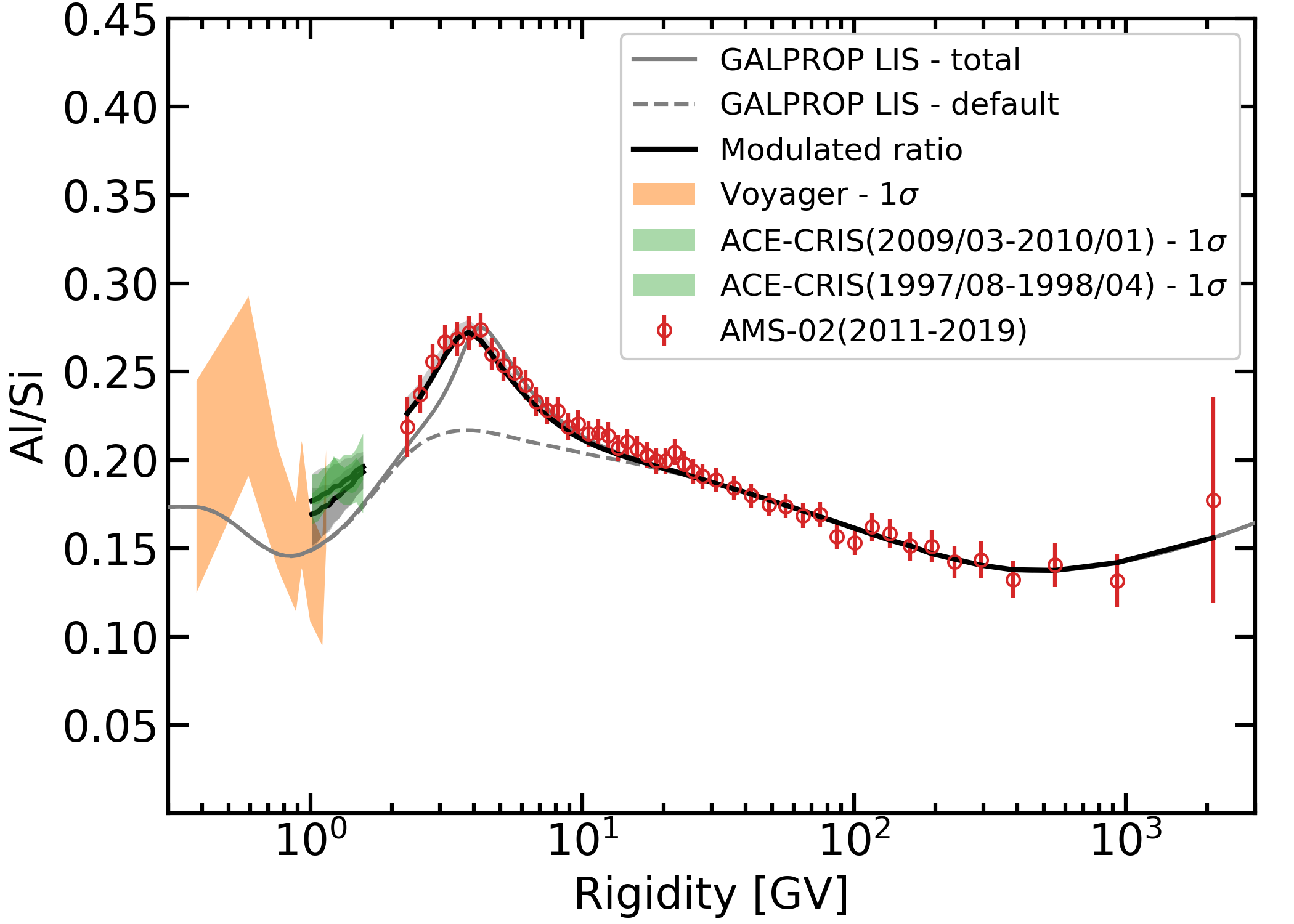}
	\caption{
The calculated Al/F and Al/Si ratios as compared with Voyager~1 \citep{2016ApJ...831...18C}, ACE-CRIS \citep{2013ApJ...770..117L}, HEAO-3-C2 \citep{1990A&A...233...96E}, and AMS-02 data \citep{2020PhRvL.124u1102A, 2021PhRvL.126h1102A, 2021PhRvL.127b1101A}. In all panels, the solid gray line shows the corresponding LIS ratio, while the modulated ratios are shown with solid black lines. The top panels show calculations for a default Al injection spectrum (Table~\ref{tbl-inject}); in the left panel shown is the ratio with the renormalized ($\times$0.896) fluorine spectrum \citep[for details see][]{2022ApJ...925..108B}. The bottom panels show calculations with an additional low-energy primary excess component on top of the default injection spectrum. For a comparison, the dashed gray lines show the default LIS ratios. The corresponding modulated ratios are shown with solid black lines. The gray shaded areas indicate 1$\sigma$ confidence limits for the calculated modulated ratios.
The Voyager 1 and ACE-CRIS data are converted from kinetic energy per nucleon to rigidity (assuming $A/Z=2$ for Si). These data are shown as shaded areas with the width corresponding to 1$\sigma$ error. 
}
	\label{fig:Al-F-Si-ratio}
\end{figure*}

\begin{figure*}[tbh!]
	\centering
	\includegraphics[width=0.49\textwidth]{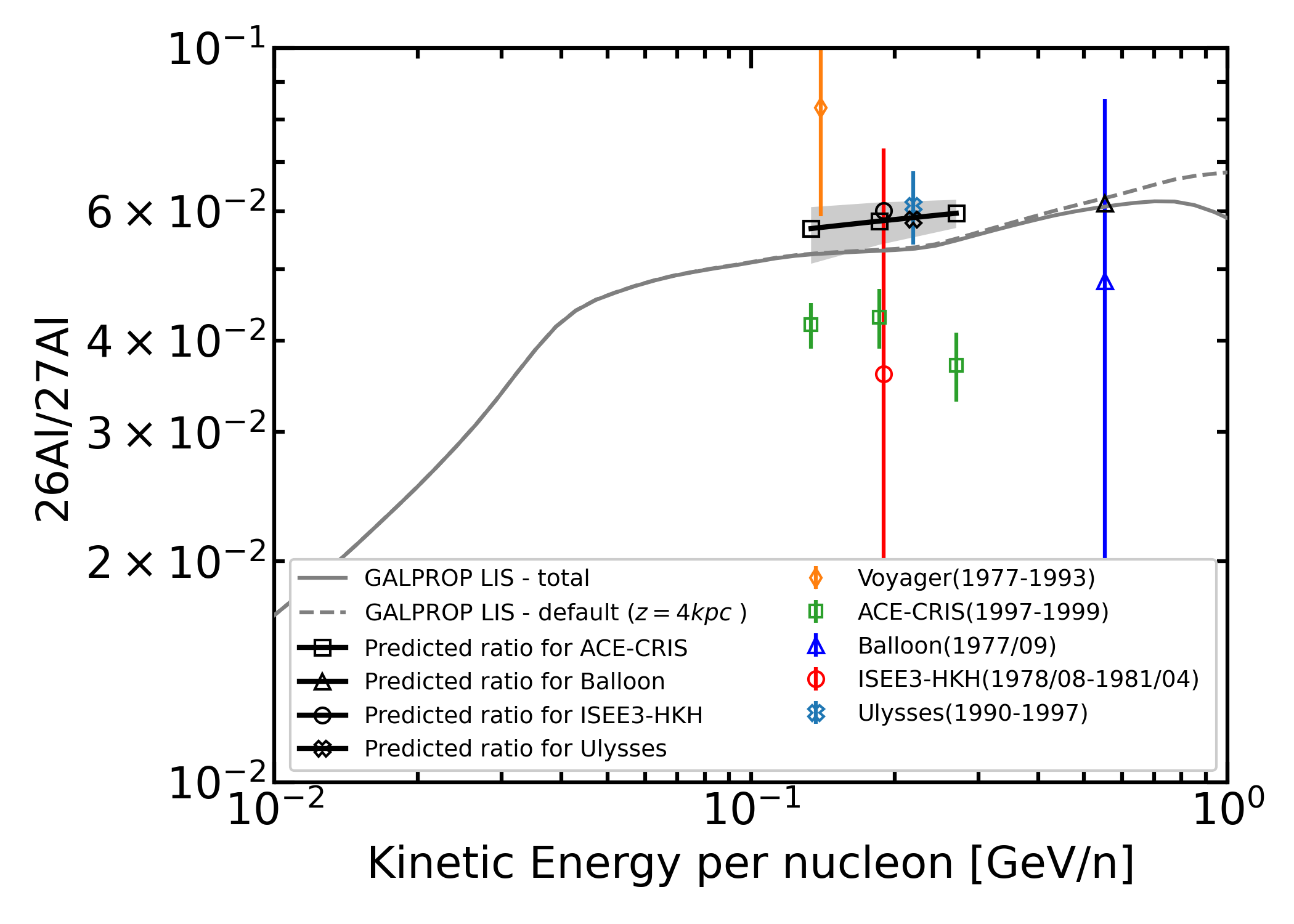}\hfill
	\includegraphics[width=0.49\textwidth]{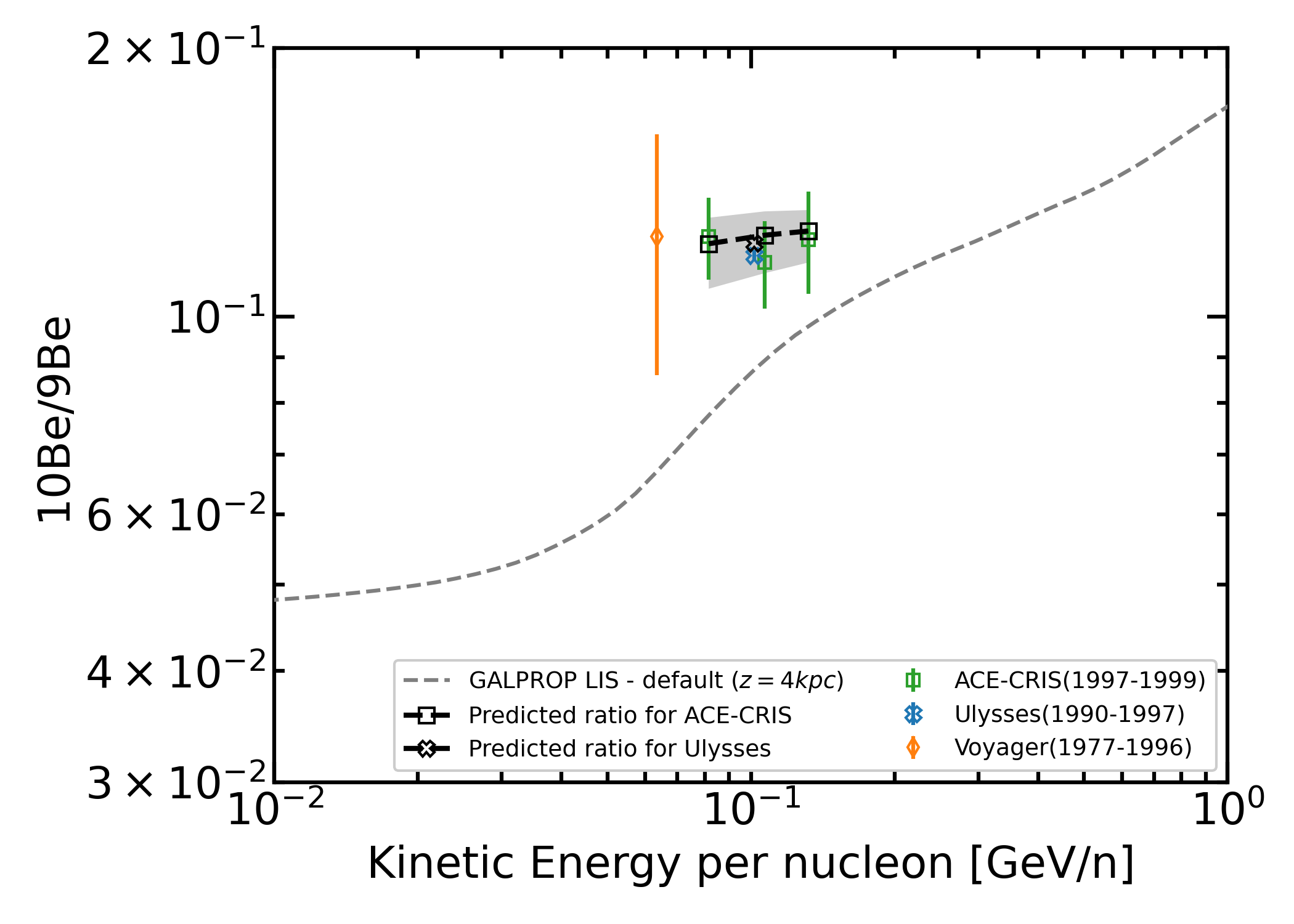}\\
	\includegraphics[width=0.49\textwidth]{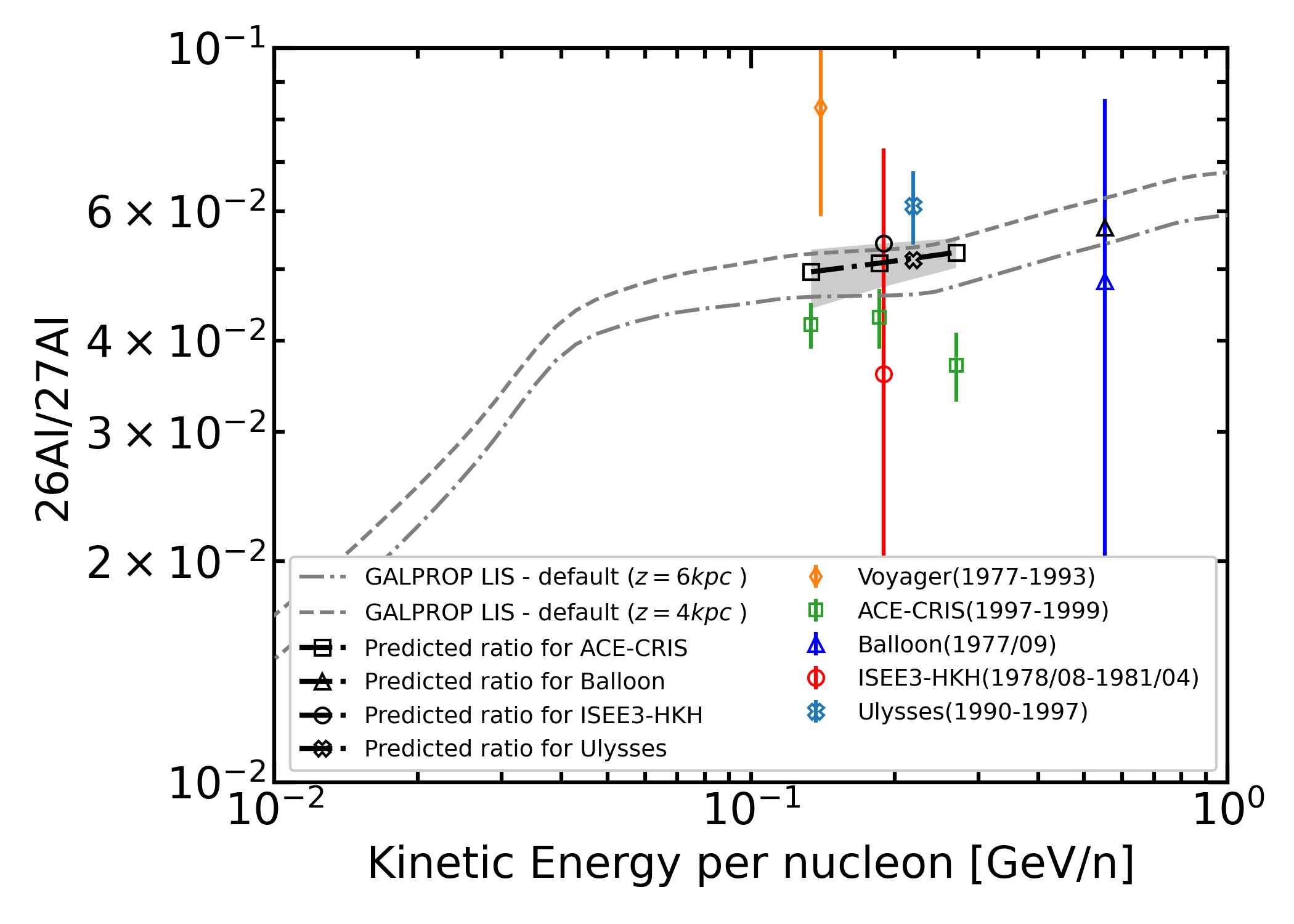}\hfill
	\includegraphics[width=0.49\textwidth]{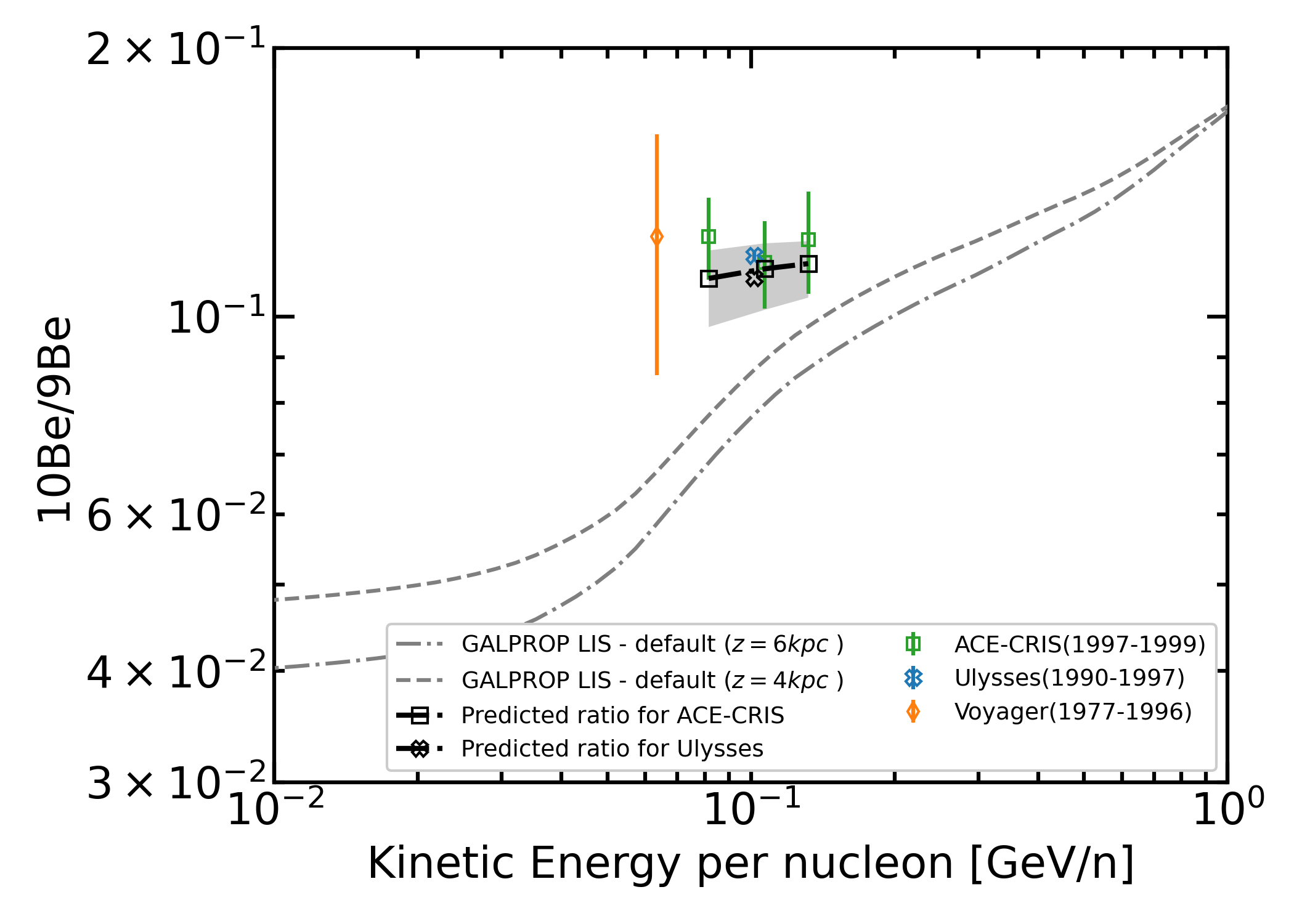}
	\caption{A comparison of the calculated isotopic ratios $^{26}$Al/$^{27}$Al (left panels) and $^{10}$Be/$^{9}$Be (right panels) with available data: Voyager 1, 2 data \citep{1994ApJ...430L..69L} collected over a period of 16 years (1977--1993) in the heliosphere, Ulysses \citep{1998ApJ...497L..85S}, ACE-CRIS \citep{Yanasak_2001}, Balloon \citep{1982ApJ...252..386W}, and ISEE3-HKH \citep{1983ICRC....9..147W}. Calculations for the halo size $z_h=4$ kpc are shown in the top panels, and for $z_h=6$ kpc in the bottom (with the adjusted normalization of the spatial diffusion coefficient $D_0 (R= 4\ {\rm GV})=7.5\times10^{28}$ cm$^{2}$ s$^{-1}$). The dashed gray lines in all plots show the default LIS ratios calculated for $z_h=4$ kpc, while the solid gray line in the top left panel shows the LIS ratio with the additional low-energy primary aluminum component. The dash-dotted lines show the default LIS ratios calculated for $z_h=6$ kpc. The colored symbols show the original published data from different instruments, while the same symbols plotted in black show our predictions based on the calculated LIS ratios for $z_h=4$ (top panels) or 6 kpc (bottom panels). The black line in the gray area shows the modulated ratio, and the gray area marks 1$\sigma$ error for our predictions.
	}
	\label{fig:Al-ratio}
\end{figure*}

Fig.~\ref{fig:Na-F-Si-ratio} shows the calculated Na/F and Na/Si ratios as compared with Voyager~1 \citep{2016ApJ...831...18C}, ACE-CRIS \citep{2013ApJ...770..117L}, and AMS-02 data \citep{2020PhRvL.124u1102A, 2021PhRvL.126h1102A, 2021PhRvL.127b1101A}. In the left panel, the dashed gray line shows the default LIS Na/F ratio (with non-renormalized fluorine spectrum), while the solid gray line shows the ratio for the renormalized ($\times$0.896) fluorine spectrum \citep[for details see][]{2022ApJ...925..108B}. The solid gray line in the right panel shows the LIS Na/Si ratio. In both panels, the modulated ratios are shown with solid black lines. The Voyager 1 and ACE-CRIS data are converted from kinetic energy per nucleon to rigidity (assuming $A/Z=2$ for Si). These data are shown as shaded areas with the width corresponding to 1$\sigma$ error. Note that the ACE-CRIS data taken during consequent solar minima overlap.

The ratios, Na/F for renormalized fluorine and Na/Si, exhibit a good overall agreement with data. 
There is some discrepancy of the calculated Na/Si ratio with AMS-02 data at 2--5 GV, which corresponds to $\approx$1$\sigma$ (5\%) if we take into account the uncertainties in the modulated ratio shown by the gray band. This can be due to the errors in the production cross sections, which we discuss in Sect.~\ref{aluminum}. At lower energies, there is a
discrepancy with the Voyager 1 data where the Na/Si ratio shows a concave shape, while the calculated ratio shows a monotonic decrease toward the lower energies. This discrepancy is related to large fluctuations (well above the experimental error bars) of Voyager 1 data points for sodium \citep[see Fig.~8 in][]{2020ApJS..250...27B}.

Fig.~\ref{fig:Al-spec} shows a comparison of the calculated Al spectra with Voyager~1 \citep{2016ApJ...831...18C}, ACE-CRIS \citep{2013ApJ...770..117L}, HEAO-3-C2 \citep{1990A&A...233...96E}, and AMS-02 data \citep{2021PhRvL.127b1101A}. The top panels show calculations for a default Al injection spectrum \citep[Table 2,][]{2020ApJS..250...27B}, which is slightly adjusted to reproduce AMS-02 data at the highest energies (Table~\ref{tbl-inject}). In the top left panel, the Al spectrum is tuned to Voyager 1,  ACE-CRIS, and the middle range of the HEAO-3-C2 instrument. The lower and higher energy HEAO-3-C2 data points are biased as was shown in \citet{2020ApJS..250...27B}, and therefore we did not not use them in our fits. To match the units of the published data, the spectra in the left panels are plotted vs.\ kinetic energy per nucleon, $E_{\rm kin}$. Obvious is a significant excess in the rigidity range 2--7 GV when compared with AMS-02 data, but outside of this range the agreement is good (Fig.~\ref{fig:Al-spec} top right).

The bottom panels show calculations with a low-energy primary excess component added to the default injection spectrum (see Table~\ref{tbl-inject}); its shape was tuned to match the AMS-02 data after propagation. The modified Al spectrum has now a clear bump, but the agreement with AMS-02 data is significantly improved at the cost of fine tuning of the injection spectrum. Possible reasons for the aluminum excess and other excesses are discussed in Sect.~\ref{aluminum}.

Fig.~\ref{fig:Al-F-Si-ratio} shows the calculated Al/F and Al/Si ratios as compared with Voyager~1 \citep{2016ApJ...831...18C} ACE-CRIS \citep{2013ApJ...770..117L}, and AMS-02 data \citep{2020PhRvL.124u1102A, 2021PhRvL.126h1102A, 2021PhRvL.127b1101A}. The Voyager 1 and ACE-CRIS data are converted from kinetic energy per nucleon to rigidity (assuming $A/Z=2$ for Si). These data are shown as shaded areas with the width corresponding to 1$\sigma$ error. The ACE-CRIS data taken during consequent solar minima are overlapping in the plots. In all panels, the solid gray line shows the corresponding LIS ratio, while the modulated ratios are shown with solid black lines. The top panels show calculations for a default Al injection spectrum (Table~\ref{tbl-inject}); in the left panel shown is the ratio with the renormalized ($\times$0.896) fluorine spectrum \citep[for details see][]{2022ApJ...925..108B}. Again, one can see a significant excess in the rigidity range 2--7 GV, while outside of this range the agreement with data is good. The bottom panels show calculations with a low-energy primary excess component  (see Table~\ref{tbl-inject} and Fig.~\ref{fig:Al-spec} bottom) added to the default injection spectrum. For a comparison, the dashed gray lines show the default LIS ratios. Again, an additional low-energy component significantly improves the agreement with data.  

The radioactive $^{26}$Al isotope in CR is entirely secondary. Therefore, it provides an independent test of the propagation model and our tuned Al spectrum. The calculated isotopic ratio $^{26}$Al/$^{27}$Al is compared with available CR data in Fig.~\ref{fig:Al-ratio}.
In the top right panel, we also show the $^{10}$Be/$^{9}$Be ratio calculated for the same propagation parameters (Table~\ref{tbl-prop}). Bottom panels show both ratios calculated for a larger halo size $z_h=6$ kpc with the adjusted normalization of the spatial diffusion coefficient $D_0 (R= 4\ {\rm GV})=7.5\times10^{28}$ cm$^{2}$ s$^{-1}$.

The calculated LIS and modulated ratios are compared with the most precise CR isotopic data available: Voyager 1, 2 data \citep{1994ApJ...430L..69L} collected over a period of 16 years (1977--1993) as the spacecraft moved from 1 au to 53 au and 41 au, correspondingly, Ulysses \citep{1998ApJ...497L..85S}, and ACE-CRIS \citep{Yanasak_2001}. In the plots the colored symbols show the original published data from different instruments, while the same symbols plotted in black show our predictions based on the calculated LIS. The gray area marks 1$\sigma$ error for our predictions. 

One can see that the agreement with Voyager 1, 2 and Ulysses data in the top left panel ($^{26}$Al/$^{27}$Al) is good, while the discrepancy with ACE-CRIS data reaches 3$\sigma$--4$\sigma$ (green squares vs.\ the shaded area with black squares). The increase in the halo size to $z_h=6$ kpc improves the agreement to $\sim$1$\sigma$ for all data sets (bottom left). The $^{10}$Be/$^{9}$Be ratio agrees well with available data in both cases $z_h=4, 6$ kpc. 

The difference in the halo size as derived from different ``radioactive clocks'' is a well-known issue \citep{2001ICRC....5.1836M, 2005AIPC..769.1612M}, where the halo size derived from the $^{10}$Be/$^{9}$Be ratio is systematically lower ($z_h=1.5-6$ kpc) than that derived from other ratios, $^{26}$Al/$^{27}$Al, $^{36}$Cl/Cl, $^{54}$Mn/Mn; meanwhile, $z_h=4-6$ kpc range was found to be consistent will all ratios. We note that calculations in \citet{2001ICRC....5.1836M, 2005AIPC..769.1612M} were done using the force-field approximation and CR data available at that time.

Overall, the agreement of the calculated $^{10}$Be/$^{9}$Be and $^{26}$Al/$^{27}$Al ratio with data is good (Fig.~\ref{fig:Al-ratio}), but indicates that halo size limits perhaps have to be reassessed based on new CR data and advances in treatment of the heliospheric modulation. On the other hand, we emphasize that the production cross sections for Be isotopes are known better than that for isotopes of Al, Cl, and Mn, and this has to be taken into account in a new assessment.

In the Appendix, we provide analytical parameterizations of the Na and default Al LIS, Eqs.~(\ref{eq:Na}), (\ref{eq:BasicAl}) and the total Al LIS, Eq.~(\ref{eq:CompleteAl}), that includes also the additional primary component. The parameters are provided in Tables \ref{Tbl-parEqNa}--\ref{Tbl-parEqCompleteAl}, correspondingly. We also provide numerical tables for all three cases, which tabulate the Na and Al LIS in rigidity $R$ (Tables \ref{Tbl-SodiumLIS-Rigi}, \ref{Tbl-BasicAluminumLIS-Rigi}) and in kinetic energy $E_{\rm kin}$ per nucleon (Tables \ref{Tbl-SodiumLIS-EKin}, \ref{Tbl-BasicAluminumLIS-EKin}), and corresponding Tables \ref{Tbl-CompleteAluminumLIS-Rigi}, \ref{Tbl-CompleteAluminumLIS-EKin} for the total Al LIS, that includes also the additional primary component.

\section{Low-energy Excesses in Lithium, Fluorine, Aluminum, and Iron}\label{aluminum}
%%%%%%%%%%%%%%%%%%%

In this Section we are speculating on the physical reasons for the observed low-energy excess in the aluminum spectrum in context of other excesses that were found in recent CR data. We do not discuss here a possibility of instrumental errors as they are rather unlikely. 

The AMS-02 detector, with its multiple redundancies, provides the state-of-the-art measurements of the CR species \citep{2021PhR...894....1A}. Before its launch, AMS-02 was tested extensively at the CERN test beam with electrons, positrons, protons, and pions. During more than 10 years of AMS-02 construction, a large international group of physicists have developed a comprehensive Monte Carlo simulation program for AMS-02. Importantly, the fragmentation cross sections of CR species inside the instrument are measured by the AMS-02 itself using the silicon tracker layers. The procedure and survival probabilities for different elements are described in \citet{2021PhR...894....1A}. The data analysis is performed in parallel by several groups, which employ different methods. A good agreement between their results is required before a paper is submitted for a publication.

Let us first discuss the default injection spectrum of Al. It is usually assumed that the majority of CR particles are accelerated in supernova remnant (SNR) shocks. The diffusive shock acceleration operating in the shocks as well as the processes in the ISM are insensitive to the nature of particles, while the relevant variable is rigidity. The standard SNR shock acceleration  \citep{2001RPPh...64..429M} results in almost identical spectra of injected particles in terms of rigidity. Therefore, the default injection spectrum of Al should be very similar to the spectra of other elements in the Si group, except perhaps at very low energies which could be sensitive to the local distributions of the ISM gas and CR sources. Table~\ref{tbl-inject} shows the default Al injection spectrum, which is similar to the injection spectra of its neighbors indeed \citep[see Table 2 in][]{2020ApJS..250...27B}. There is no viable reason of why the Al injection spectrum in the 2--10 GV range should be different from its neighbors. This forces us to the conclusion that the observed excess must have a different origin, which is discussed below.

There are four possible physical reasons of the observed excess in the Al spectrum: (i) an incorrect spectrum of $^{28}$Si, the major progenitor of secondary $^{26,27}$Al, (ii) errors in the total inelastic cross sections of Al, (iii) errors in the isotopic production cross sections of $^{26,27}$Al, and (iv) an additional local component of primary Al. Below we are discussing them one by one. For completeness, in point (iii) we are also discussing the accuracy of $^{23}$Na cross sections. The calculated spectra of Mg and Si, the major progenitors of Na, are tuned to the recent AMS-02 data. 

(i) The Si spectrum is tuned to Voyager 1, ACE-CRIS, and AMS-02 data \citep{2020ApJS..250...27B}. Importantly, AMS-02 data are available above 2 GV, the same rigidity range for all CR species. This is exactly the energy range where the Al excess is observed. Contributions of other CR species are very minor and cannot be responsible for the observed excess. 

(ii) Significant errors in the total inelastic cross section of Al as a primary reason for the excess can be excluded by consideration of the accelerator data. \citet{Bobchenko:1979hp} provide a table of measured proton-nucleus cross sections for the proton momenta from 5 GeV/c to 9 GeV/c that correspond to the ambient $^{27}$Al rigidity range 10--19 GV in the inverse kinematics, while the measurement accuracy is stated as 1-2\%. The average value of the total inelastic cross section in this interval is 456.1 mb and typical error bars are $\pm$4 mb. Most of the data points are consistent with a constant value within the error bars with just four out of total seventeen points being clear outliers at 443$\pm$4 mb, 445$\pm$4 mb, and 467$\pm$4 mb (two points). This is the energy range where the total inelastic cross sections are flat \citep[e.g., see Fig.~1 in][]{1996PhRvC..54.1329W} and they are easier to measure than the isotopic production. There are also abundant data on the total inelastic cross section of $^{27}$Al below $\sim$1 GeV nucleon$^{-1}$ ($<$3 GV) \citep{BAUHOFF1986429}. The parameterizations of the total inelastic cross sections used in our calculations \citep[e.g.,][]{BarPol1994} are tuned to the available data. That makes a significant error below 7 GV, where the excess is observed, rather unlikely. We note the absence of a similar excess in the spectra of neighboring nuclei, such as Ne, Mg, Si \citep{2020ApJS..250...27B}.

(iii) Almost 100\% of secondary $^{27}$Al is produced through fragmentation of $^{28}$Si with some minor contributions from $^{29}$Si, $^{32}$S, and $^{56}$Fe \citep{2013ICRC...33..803M}. Unfortunately, the isotopic production cross section $p+ ^{28}$Si$\to$$^{27}$Al is a major source of uncertainty. There are only a couple of measurements available for this reaction:  $(E_{\rm kin},\sigma)$ = (218 MeV nucleon$^{-1}$, 25 mb) \citep{1975PhRvL..35..773A}, and (506 MeV nucleon$^{-1}$, 48.5 mb) \citep{1998PhRvC..58.3539W}. These energies correspond to 1.4 GV and 2.3 GV ($^{27}$Al) in the inverse kinematics, respectively, and constrain the production of $^{27}$Al in the lower range of rigidities where the excess is observed. Clearly, the full energy dependence of the cross section can only be evaluated using semi-phenomenological calculations or nuclear codes \citep[see the description of \galprop{} in][]{2020ApJS..250...27B}. However, measurements of $^{28}$Si fragmentation $^{28}$Si$+p\to$Al made by a Siegen group \citep{1991ICRC....2..280H} indicate that the cross section remains flat between 1 and 14.5 GeV nucleon$^{-1}$. This energy range corresponds to the rigidity range from 3.5--31 GV ($^{27}$Al), which covers the excess range and extends to significantly higher rigidities. Therefore, even though the cross section between $\approx$500 MeV nucleon$^{-1}$ and $\approx$1000 MeV nucleon$^{-1}$ remains unconstrained, it has no effect on the excess, which is observed in the rigidity range 2--7 GV with the maximum at $\approx$4 GV. The latter corresponds to $E_{\rm kin}\approx$ 1.2 GeV nucleon$^{-1}$, i.e.\ where the cross section is flat.

To further investigate the issue, we look at another aluminum isotope, $^{26}$Al, which is mostly produced through fragmentation of $^{28}$Si and $^{27}$Al \citep{2013ICRC...33..803M}. Since we compare our calculations with the observed CR isotopic ratio $^{26}$Al/$^{27}$Al below 1 GeV nucleon$^{-1}$ (Fig.~\ref{fig:Al-ratio}), our concern is the accuracy of calculations of production of $^{26}$Al in this energy range with a correction for heliospheric modulation. Fortunately, production of $^{26}$Al is well-constrained below 1 GeV nucleon$^{-1}$. The experimental data points with references can be found in the EXFOR\footnote{https://www-nds.iaea.org/exfor/} data base; the data points from different experiments are consistent within the error bars.

The calculated $^{26}$Al/$^{27}$Al ratio (Fig.~\ref{fig:Al-ratio}, bottom left, $z_h=6$ kpc) agrees well with available CR data corrected for the corresponding level of heliospheric modulation. To summarize, it is unlikely that the uncertainties in the production cross sections of $^{26,27}$Al can result in the observed excess.

We also comment here on the accuracy of sodium cross sections. The total inelastic cross section is calculated using the same parameterizations that are used for other species \citep{BarPol1994}. It should be accurate enough given that total inelastic cross sections of its neighbors, Ne and Mg, are well-measured. The production cross sections of $^{23}$Na are less certain. The secondary sodium $^{23}$Na is produced through fragmentation of $^{24}$Mg (dominant channel) and $^{28}$Si with minor contributions of $^{25}$Mg, $^{26}$Mg, and $^{27}$Al \citep{2013ICRC...33..803M}. Unfortunately, there is only a single measurement for each parent nucleus, $^{24,25,26}$Mg, $^{28}$Si, $^{27}$Al, made by \citet{1990PhRvC..41..547W} at the LBL Bevalac using a $\sim$600 MeV nucleon$^{-1}$ beams. Therefore, we are using the updated semi-phenomenological parameterizations by \citet{2003ApJS..144..153W} renormalized to the data. The beam energy of $\sim$600 MeV nucleon$^{-1}$ (corresponds to $^{23}$Na rigidity of $\sim$2.5 GV) above which the expected cross section behavior is quite smooth. It is reasonable to expect that Na production cross sections should be flat above 2.5 GV, i.e.\ extending into the rigidity range  2--7 GV of the observed Al excess.

(iv) The observed low-energy excess in aluminum spectrum is most likely due to the stable isotope $^{27}$Al. However, because it is stable, its yield and Galactic distribution are difficult to measure. Meanwhile, abundant literature exists on the observations of the distribution of the diffuse \gray{} 1.809 MeV line emission from the decay of the radioactive $^{26}$Al isotope and on its origin. Therefore, we are discussing the origin of $^{26}$Al assuming that the same sources are producing $^{27}$Al as well. 

Studies of the primitive meteoritic grains indicate that the early solar system material was enriched with short-lived radionuclides, such as $^{10}$Be, $^{26}$Al, $^{36}$Cl, $^{41}$Ca, $^{53}$Mn, $^{60}$Fe, and others \citep{2005ASPC..341..485G}. In particular, the presolar meteoritic ratio $^{26}$Al/$^{27}$Al$\sim$$4.5\times10^{-5}$ derived from the excess of $^{26}$Mg is surprisingly well determined despite the fact that the half-life of $^{26}$Al of $\sim$700 kyr is much shorter than the formation time of the solar system. Therefore, a local source of $^{26}$Al, such as nearby SN or an AGB star, which injected fresh $^{26}$Al into the forming solar system, is usually invoked.

Observations of the distribution of the diffuse Galactic 1.809 MeV \gray{} emission line from $^{26}$Al decay by COMPTEL \citep{1995A&A...298..445D, 1996A&AS..120C.311O, 1999A&A...345..813K} and INTEGRAL \citep{2015ApJ...801..142B} have shown that $^{26}$Al nucleosynthesis is ongoing in the present Galaxy. Potential sources include AGB stars, novae, core collapse supernovae (SNe), and Wolf-Rayet (WR) star winds \citep{1996PhR...267....1P, 2006Natur.439...45D}. Very recent review of the nucleosynthesis reactions involved in the production and destruction of $^{26}$Al and $^{60}$Fe and characteristics of the stellar sites of their nucleosynthesis can be found in \citet{2021arXiv210908558D}.

Apparently, the sources of cosmic Al are numerous, and are simultaneously also the sources of other rare isotopes, such as $^7$Li, $^{19}$F, $^{60}$Fe and others \citep[see discussions in our papers:][and references therein]{2020ApJ...889..167B, 2021ApJ...913....5B, 2022ApJ...925..108B}. Particularly interesting is a possible contribution of the massive WR stars proposed to explain the observed anomalous $^{22}$Ne/$^{20}$Ne ratio and other observed ratios, $^{12}$C/$^{16}$O, and $^{58}$Fe/$^{56}$Fe, in CRs \citep{2008NewAR..52..427B}. It was shown that all isotopic ratios measured with ACE-CRIS are consistent with CR source composition consisting of $\sim$20\% of WR material mixed with $\sim$80\% material with solar-system composition. Since most of WR stars are found in the OB associations and are the evolutionary stages of OB stars, the OB associations (through multiple SN explosions) are the likely sources of a substantial fraction of CRs. WR stars are also most likely local sources of $^7$Li, and $^{19}$F, while $^{60}$Fe may come from a SN or a series of SN explosions 1--2 Myr ago in the solar neighborhood \citep{2016Sci...352..677B}.

Regarding the spectrum of sodium, we note the absence of a similar low-energy excess. An interesting analysis using the Gaia-ESO Survey to study sodium and aluminum abundances in giants and dwarfs and its implications for stellar and Galactic chemical evolution can be found in \citet{2016A&A...589A.115S}. WR stars do not seem to be a significant source of sodium. Absence of the excess in the sodium spectrum apparently supports the hypothesis of the origin of the observed excesses in the local OB associations. Meanwhile, studies of sodium nucleosynthesis in different stellar environments are desirable as they may help to discriminate between possible scenarios of the origin of the observed excesses.  

\section{Conclusion} \label{conclusion}
%%%%%%%%%%%%%%%%%%%%%%%%%%%%%%%%%%%%%%%%%%%%%%%%%%%
%%%%%%%%%%%%%%%%%%%%%%%%%%%%%%%%%%%%%%%%%%%%%%%%%%%

Using the combined data of AMS-02 \citep{2021PhRvL.126d1104A}, ACE-CRIS \citep{2013ApJ...770..117L}, and Voyager 1 \citep{2016ApJ...831...18C} we analyzed the spectra of sodium and aluminum over a wide rigidity range from 100 MV to 2 TV. We found that the sodium spectrum agrees well with the predictions made with the \galprop{}-\helmod{} framework, while aluminum spectrum shows a significant excess in the rigidity range from 2--7 GV. 

Our analysis shows that among four possible reasons of the observed excess, a contribution of local sources is most likely. In this context, we discuss the origin of other low-energy excesses in Li, F, and Fe found earlier \citep{2020ApJ...889..167B, 2021ApJ...913....5B, 2022ApJ...925..108B}. It seems that the WR hypothesis that was invoked to reproduce the observed isotopic ratios, $^{22}$Ne/$^{20}$Ne, $^{12}$C/$^{16}$O, $^{58}$Fe/$^{56}$Fe, and other observed CR abundances \citep{2008NewAR..52..427B} is also consistent with the observed excesses in Li, F, and Al, while excess in Fe may be connected with a past SN activity in the solar neighborhood. Absence of a corresponding excess in sodium supports this hypothesis, as the WR winds are not a significant source of sodium.

Finally, we note that the exploration of the low-energy features in the spectra of CR species has just begun, thanks to the data from the interstellar probes Voyager 1, 2, and precise measurements by AMS-02, ACE-CRIS, and other instruments. These features harbor the keys to understanding our local Galactic environment and the history of formation of the solar system. The increase in the collected statistics and reduction in the systematic errors will help to establish the precise spectral shapes of observed features and to facilitate their interpretation. 

\acknowledgements
Special thanks to Pavol Bobik, Giuliano Boella, Marian Putis, and Mario Zannoni for their continuous support of the \helmod{} project and many useful suggestions. 
This work was carried out using \helmod{} tool, which is currently supported within ASIF--ASI (Agenzia Spaziale Italiana) Supported Irradiation Facilities--framework for space radiation environment activities, e.g., ASIF implementation agreements 2017-22-HD.0 ASI-ENEA, 2017-15-HD.0 ASI-INFN, 2021-39-HH.0 ASI-ENEA and ASIF implementation agreement 2021-36-HH.0 involving ASI and Milano-Bicocca University. This work is also supported by ASI contract ASI-INFN 2019-19-HH.0 and ESA (European Space Agency) contract 4000116146/16/NL/HK.
Igor V.\ Moskalenko and Troy A.\ Porter acknowledge support from NASA Grants No.~NNX17AB48G, 80NSSC22K0718, 80NSSC22K0477. This research has made use of the SSDC Cosmic rays database \citep{2017ICRC...35.1073D} and LPSC Database of Charged Cosmic Rays \citep{2014A&A...569A..32M}.

\bibliography{bibliography}

%\appendix
%\include{NaAl_Tables_formulas}

\appendix
%\allowdisplaybreaks

\section{Analytical parameterization and numerical tables of the sodium and aluminum LIS}

Here we provide an analytical parameterization of the sodium LIS (m$^{-2}$ sr$^{-1}$ s$^{-1}$ GV$^{-1}$): 
\begin{align}
	% -------------------------------------------------- Z = 11 Sodium ---------------------------------------------
	\label{eq:Na}
	F_{\rm Na} (R)  = %\\
	&\begin{cases}
		\displaystyle a - b R\tilde{R}  - G(R)\left( c + dR - f\,R^3 \right)  - g\left(h\,R^4\right)^{G(R)},    &R\le 2.8\, {\rm GV}, \smallskip\\
		\displaystyle R^{-2.7} \left[ - i - lR + m\sqrt{n - \frac{o}{R} + p\sqrt{R}} + qR\tanh\left(-\frac{rR}{sR + R^2 - t\sqrt{R}} \right)  \right],  &R> 2.8\, {\rm GV},
	\end{cases}%\\[8pt] 
\end{align}
where $R$ is the particle rigidity in GV, expressions for $\tilde{R}$ and $G(x)$ are given in Eq.~(\ref{RGL}), and the values of the fitting parameters from $a$ to $z$ are summarized in Table \ref{Tbl-parEqNa}.

\begin{deluxetable}{rlrlrlrl}[hp]
	\tablecolumns{8}
	\tablewidth{0mm}
	%   \def\arraystretch{0.9}
	%   \tabletypesize{\scriptsize}
	\tablecaption{Parameters of the analytical fit to the sodium LIS, Eq.~\eqref{eq:Na} \label{Tbl-parEqNa}}
	\tablehead{
		\colhead{Param}  &  \colhead{Value}  & \colhead{Param}  & \colhead{Value}  & \colhead{Param}  &  \colhead{Value}  & \colhead{Param}    &  \colhead{Value}
	}
	\startdata
	$a$\phantom{a} & 2.8359e-1 & $g$\phantom{a}  & 9.2254e-2 & $n$\phantom{a} & 3.3852e+0 & $s$\phantom{a} & 4.100e+1 \\
	$b$\phantom{a} & 3.1660e-2 & $h$\phantom{a}  & 1.0122e-1 & $o$\phantom{a} & 4.0058e+0 & $t$\phantom{a} & 4.8223e+1 \\
	$c$\phantom{a} & 2.8544e-1 & $i$\phantom{a}  & 2.9718e+0 & $p$\phantom{a} & 9.9286e-4 & \nodata & \nodata \\
	$d$\phantom{a} & 7.4636e-2 & $l$\phantom{a}  & 3.3525e-7 & $q$\phantom{a} & 2.8346e-2 & \nodata & \nodata \\
	$f$\phantom{a} & 1.2551e-1 & $m$\phantom{a}  & 3.3510e+0 & $r$\phantom{a} & 7.3991e+1 & \nodata & \nodata \\
	\enddata
\end{deluxetable}

We also provide analytical parameterizations of the ``default'' aluminum LIS and total aluminum LIS (with the excess), units are m$^{-2}$ sr$^{-1}$ s$^{-1}$ GV$^{-1}$: 
\begin{align}
	% -------------------------------------------------- Z = 13 Aluminum ---------------------------------------------
	\label{eq:BasicAl}
	F_{\rm Al}^{\rm default} (R)  = %\\
	&\begin{cases}
	    \displaystyle a - \frac{b}{R} + cR^3 + dR^2 G(R) - fR^2 - g\,R\,G(R) - hG^4(R) ,    &R\le 3.6\, {\rm GV}, \smallskip\\
		\displaystyle R^{-2.7} \left[     i\tilde{R} + l\sqrt{R} + \frac{m}{\log(n + oR)} + p\log^2(n + oR) - q - r\,R\tilde{R} - s\tilde{R}^2)
\right],  &R> 3.6\, {\rm GV},
		%\nonumber
	\end{cases}%\\[8pt] 
\end{align}
\begin{align}
	% -------------------------------------------------- Z = 13 Aluminum ---------------------------------------------
	\label{eq:CompleteAl}
	F_{\rm Al}^{\rm total} (R)  = %\\
	&\begin{cases}
		\displaystyle aR + \frac{b}{R} + cR^3 + d\,R\tilde{R}^4G^2(R) - f - gR^2 - h\tilde{R}^2 ,    &R\le 3.2\, {\rm GV}, \smallskip\\
		\displaystyle R^{-2.7} \left\{ iR - l + m\tanh{(R)} + \tilde{R} \left( n - oR\right) + p\log(q + R\tilde{R}) - r\,L\left[s\log(q + R\tilde{R}) \right] \right\},  &R> 3.2\, {\rm GV},
		%\nonumber
	\end{cases}%\\[8pt] 
\end{align}
where $R$ is the particle rigidity in GV, 
\begin{align} \label{RGL}
\tilde{R}&=\log{R},\nonumber \\
G(x)&=e^{-x^2},\\
L(x)&=1/(1+e^{-x}),\nonumber %\frac{1}{1+e^{-x}},\nonumber
\end{align}
and the values of the fitting parameters from $a$ to $z$ are given in Tables \ref{Tbl-parEqBasicAl} and \ref{Tbl-parEqCompleteAl}.

\begin{deluxetable}{rlrlrlrl}[hp]
	\tablecolumns{8}
	\tablewidth{0mm}
	%   \def\arraystretch{0.9}
	%   \tabletypesize{\scriptsize}
	\tablecaption{Parameters of the analytical fit to the ``default'' aluminum LIS, Eq.~\eqref{eq:BasicAl} \label{Tbl-parEqBasicAl}}
	\tablehead{
		\colhead{Param}  &  \colhead{Value}  & \colhead{Param}  & \colhead{Value}  & \colhead{Param}  &  \colhead{Value}  & \colhead{Param}    &  \colhead{Value}
	}
	\startdata
	$a$\phantom{a}   & 2.1141e-1 & $g$\phantom{a}  & 2.0344e-1 & $n$\phantom{a} & 2.7104e+0 & $s$\phantom{a} & 1.4698e-1\\
	$b$\phantom{a}   & 9.7044e-4 & $h$\phantom{a}  & 1.8791e-1 & $o$\phantom{a} & 3.0751e-2 & \nodata & \nodata\\
	$c$\phantom{a}   & 4.0776e-3 & $i$\phantom{a}  & 1.8105e+0 & $p$\phantom{a} & 1.4367e-1 & \nodata & \nodata\\
	$d$\phantom{a}   & 1.0513e-1 & $l$\phantom{a}  & 2.2979e-2 & $q$\phantom{a} & 7.6313e+0 & \nodata & \nodata\\
	$f$\phantom{a}   & 2.6449e-2 & $m$\phantom{a}  & 7.4947e+0 & $r$\phantom{a} & 3.1067e-7 & \nodata & \nodata\\
	\enddata
\end{deluxetable}

\begin{deluxetable}{rlrlrlrl}[!hp]
	\tablecolumns{8}
	\tablewidth{0mm}
	%   \def\arraystretch{0.9}
	%   \tabletypesize{\scriptsize}
	\tablecaption{Parameters of the analytical fit to the total aluminum LIS, Eq.~\eqref{eq:CompleteAl} \label{Tbl-parEqCompleteAl}}
	\tablehead{
		\colhead{Param}  &  \colhead{Value}  & \colhead{Param}  & \colhead{Value}  & \colhead{Param}  &  \colhead{Value}  & \colhead{Param}    &  \colhead{Value}
	}
	\startdata
	$a$\phantom{a} & 2.3066e-1 & $g$\phantom{a}  & 8.9912e-2 & $n$\phantom{a} & 6.3989e-1 & $s$\phantom{a} & 2.3391e-1 \\
	$b$\phantom{a} & 2.4724e-2 & $h$\phantom{a}  & 9.5355e-2 & $o$\phantom{a} & 8.5007e-6 & \nodata & \nodata \\
	$c$\phantom{a} & 1.2692e-2 & $i$\phantom{a}  & 1.2796e-4 & $p$\phantom{a} & 1.6380e+0 & \nodata & \nodata \\
	$d$\phantom{a} & 9.6944e-2 & $l$\phantom{a}  & 9.5849e+1 & $q$\phantom{a} & 1.2755e+2 & \nodata & \nodata \\
	$f$\phantom{a} & 2.9140e-2 & $m$\phantom{a}  & 1.4909e+2 & $r$\phantom{a} & 7.8534e+1 & \nodata & \nodata \\
	\enddata
\end{deluxetable}

The analytical representations, Eqs.~(\ref{eq:Na})--(\ref{eq:CompleteAl}), are also complemented by numerical tables calculated for the {\it I}-scenario, which tabulate the LIS in rigidity $R$ (Tables \ref{Tbl-SodiumLIS-Rigi}, \ref{Tbl-BasicAluminumLIS-Rigi}, \ref{Tbl-CompleteAluminumLIS-Rigi}), and in kinetic energy $E_{\rm kin}$ per nucleon (Tables \ref{Tbl-SodiumLIS-EKin}, \ref{Tbl-BasicAluminumLIS-EKin}, \ref{Tbl-CompleteAluminumLIS-EKin}).

\begin{deluxetable*}{cccccccccc}%[p!]
	\tabletypesize{\footnotesize}
	\tablecolumns{10}
	\tablewidth{0mm}
	\tablecaption{$Z=11$ -- Sodium LIS\label{Tbl-SodiumLIS-Rigi}}
	\tablehead{
		\colhead{Rigidity} & \colhead{Differential} &
		\colhead{Rigidity} & \colhead{Differential} &
		\colhead{Rigidity} & \colhead{Differential} &
		\colhead{Rigidity} & \colhead{Differential} &
		\colhead{Rigidity} & \colhead{Differential} \\ [-2ex] 
		\colhead{GV} & \colhead{intensity} &
		\colhead{GV} & \colhead{intensity} &
		\colhead{GV} & \colhead{intensity} &
		\colhead{GV} & \colhead{intensity} &
		\colhead{GV} & \colhead{intensity}  
	}
	\startdata
9.027e-02 & 7.982e-04 & 7.289e-01 & 1.083e-01 & 1.008e+01 & 4.958e-03 & 5.271e+02 & 5.598e-08 & 3.313e+04 & 7.803e-13\\
9.473e-02 & 9.152e-04 & 7.661e-01 & 1.167e-01 & 1.094e+01 & 3.994e-03 & 5.802e+02 & 4.277e-08 & 3.649e+04 & 6.047e-13\\
9.941e-02 & 1.035e-03 & 8.054e-01 & 1.251e-01 & 1.188e+01 & 3.206e-03 & 6.387e+02 & 3.271e-08 & 4.018e+04 & 4.687e-13\\
1.043e-01 & 1.171e-03 & 8.469e-01 & 1.334e-01 & 1.292e+01 & 2.563e-03 & 7.032e+02 & 2.504e-08 & 4.424e+04 & 3.633e-13\\
1.095e-01 & 1.324e-03 & 8.906e-01 & 1.413e-01 & 1.405e+01 & 2.040e-03 & 7.741e+02 & 1.919e-08 & 4.872e+04 & 2.817e-13\\
1.149e-01 & 1.498e-03 & 9.369e-01 & 1.488e-01 & 1.530e+01 & 1.617e-03 & 8.523e+02 & 1.471e-08 & 5.365e+04 & 2.185e-13\\
1.206e-01 & 1.694e-03 & 9.857e-01 & 1.556e-01 & 1.668e+01 & 1.277e-03 & 9.383e+02 & 1.129e-08 & 5.908e+04 & 1.695e-13\\
1.265e-01 & 1.916e-03 & 1.037e+00 & 1.619e-01 & 1.819e+01 & 1.004e-03 & 1.033e+03 & 8.668e-09 & 6.506e+04 & 1.315e-13\\
1.328e-01 & 2.167e-03 & 1.092e+00 & 1.674e-01 & 1.985e+01 & 7.862e-04 & 1.137e+03 & 6.661e-09 & 7.164e+04 & 1.020e-13\\
1.393e-01 & 2.451e-03 & 1.150e+00 & 1.721e-01 & 2.168e+01 & 6.136e-04 & 1.252e+03 & 5.122e-09 & 7.889e+04 & 7.918e-14\\
1.462e-01 & 2.772e-03 & 1.211e+00 & 1.760e-01 & 2.369e+01 & 4.771e-04 & 1.379e+03 & 3.939e-09 & 8.687e+04 & 6.146e-14\\
1.535e-01 & 3.134e-03 & 1.276e+00 & 1.789e-01 & 2.590e+01 & 3.696e-04 & 1.518e+03 & 3.031e-09 & 9.566e+04 & 4.772e-14\\
1.611e-01 & 3.543e-03 & 1.345e+00 & 1.809e-01 & 2.834e+01 & 2.854e-04 & 1.671e+03 & 2.333e-09 & 1.053e+05 & 3.705e-14\\
1.690e-01 & 4.005e-03 & 1.419e+00 & 1.816e-01 & 3.103e+01 & 2.199e-04 & 1.840e+03 & 1.796e-09 & 1.160e+05 & 2.878e-14\\
1.774e-01 & 4.526e-03 & 1.497e+00 & 1.812e-01 & 3.398e+01 & 1.690e-04 & 2.026e+03 & 1.383e-09 & 1.277e+05 & 2.235e-14\\
1.862e-01 & 5.114e-03 & 1.580e+00 & 1.795e-01 & 3.723e+01 & 1.295e-04 & 2.231e+03 & 1.066e-09 & 1.407e+05 & 1.736e-14\\
1.954e-01 & 5.776e-03 & 1.668e+00 & 1.764e-01 & 4.081e+01 & 9.914e-05 & 2.457e+03 & 8.214e-10 & 1.549e+05 & 1.349e-14\\
2.050e-01 & 6.522e-03 & 1.763e+00 & 1.720e-01 & 4.475e+01 & 7.572e-05 & 2.705e+03 & 6.332e-10 & 1.706e+05 & 1.049e-14\\
2.152e-01 & 7.362e-03 & 1.863e+00 & 1.662e-01 & 4.909e+01 & 5.773e-05 & 2.979e+03 & 4.883e-10 & 1.878e+05 & 8.150e-15\\
2.259e-01 & 8.307e-03 & 1.971e+00 & 1.592e-01 & 5.387e+01 & 4.394e-05 & 3.280e+03 & 3.766e-10 & 2.068e+05 & 6.335e-15\\
2.371e-01 & 9.369e-03 & 2.087e+00 & 1.510e-01 & 5.913e+01 & 3.340e-05 & 3.612e+03 & 2.905e-10 & \nodata & \nodata\\
2.488e-01 & 1.056e-02 & 2.211e+00 & 1.419e-01 & 6.492e+01 & 2.536e-05 & 3.977e+03 & 2.241e-10 & \nodata & \nodata\\
2.611e-01 & 1.190e-02 & 2.344e+00 & 1.321e-01 & 7.130e+01 & 1.923e-05 & 4.379e+03 & 1.729e-10 & \nodata & \nodata\\
2.741e-01 & 1.340e-02 & 2.487e+00 & 1.220e-01 & 7.832e+01 & 1.457e-05 & 4.822e+03 & 1.334e-10 & \nodata & \nodata\\
2.877e-01 & 1.508e-02 & 2.640e+00 & 1.115e-01 & 8.605e+01 & 1.103e-05 & 5.310e+03 & 1.030e-10 & \nodata & \nodata\\
3.020e-01 & 1.695e-02 & 2.806e+00 & 1.010e-01 & 9.457e+01 & 8.347e-06 & 5.847e+03 & 7.952e-11 & \nodata & \nodata\\
3.170e-01 & 1.905e-02 & 2.985e+00 & 9.058e-02 & 1.039e+02 & 6.310e-06 & 6.438e+03 & 6.140e-11 & \nodata & \nodata\\
3.328e-01 & 2.138e-02 & 3.179e+00 & 8.052e-02 & 1.143e+02 & 4.766e-06 & 7.090e+03 & 4.742e-11 & \nodata & \nodata\\
3.493e-01 & 2.397e-02 & 3.388e+00 & 7.097e-02 & 1.256e+02 & 3.598e-06 & 7.807e+03 & 3.663e-11 & \nodata & \nodata\\
3.667e-01 & 2.686e-02 & 3.615e+00 & 6.205e-02 & 1.382e+02 & 2.716e-06 & 8.596e+03 & 2.830e-11 & \nodata & \nodata\\
3.850e-01 & 3.005e-02 & 3.862e+00 & 5.379e-02 & 1.519e+02 & 2.050e-06 & 9.466e+03 & 2.187e-11 & \nodata & \nodata\\
4.042e-01 & 3.358e-02 & 4.129e+00 & 4.627e-02 & 1.671e+02 & 1.547e-06 & 1.042e+04 & 1.690e-11 & \nodata & \nodata\\
4.244e-01 & 3.748e-02 & 4.421e+00 & 3.951e-02 & 1.838e+02 & 1.168e-06 & 1.148e+04 & 1.307e-11 & \nodata & \nodata\\
4.456e-01 & 4.178e-02 & 4.738e+00 & 3.352e-02 & 2.022e+02 & 8.820e-07 & 1.264e+04 & 1.010e-11 & \nodata & \nodata\\
4.679e-01 & 4.649e-02 & 5.083e+00 & 2.831e-02 & 2.225e+02 & 6.664e-07 & 1.392e+04 & 7.814e-12 & \nodata & \nodata\\
4.913e-01 & 5.163e-02 & 5.460e+00 & 2.377e-02 & 2.448e+02 & 5.039e-07 & 1.533e+04 & 6.045e-12 & \nodata & \nodata\\
5.160e-01 & 5.724e-02 & 5.872e+00 & 1.987e-02 & 2.694e+02 & 3.813e-07 & 1.688e+04 & 4.676e-12 & \nodata & \nodata\\
5.419e-01 & 6.330e-02 & 6.323e+00 & 1.654e-02 & 2.965e+02 & 2.888e-07 & 1.858e+04 & 3.619e-12 & \nodata & \nodata\\
	\enddata
	\tablecomments{Differential Intensity units: (m$^2$ s sr GV)$^{-1}$.}
\end{deluxetable*}

\begin{deluxetable*}{cccccccccc}%[p!]
	\tabletypesize{\footnotesize}
	\tablecolumns{10}
	\tablewidth{0mm}
	\tablecaption{$Z=11$ -- Sodium LIS\label{Tbl-SodiumLIS-EKin}}
	\tablehead{
		\colhead{$E_{\rm kin}$} & \colhead{Differential} &
		\colhead{$E_{\rm kin}$} & \colhead{Differential} &
		\colhead{$E_{\rm kin}$} & \colhead{Differential} &
		\colhead{$E_{\rm kin}$} & \colhead{Differential} &
		\colhead{$E_{\rm kin}$} & \colhead{Differential} \\ [-2ex] 
		\colhead{GeV/n} & \colhead{intensity} &
		\colhead{GeV/n} & \colhead{intensity} &
		\colhead{GeV/n} & \colhead{intensity} &
		\colhead{GeV/n} & \colhead{intensity} &
		\colhead{GeV/n} & \colhead{intensity}  
	}
	\startdata
1.000e-03 & 3.605e-02 & 6.309e-02 & 6.463e-01 & 3.981e+00 & 1.056e-02 & 2.512e+02 & 1.170e-07 & 1.585e+04 & 1.632e-12\\
1.101e-03 & 3.939e-02 & 6.948e-02 & 6.668e-01 & 4.384e+00 & 8.482e-03 & 2.766e+02 & 8.942e-08 & 1.745e+04 & 1.264e-12\\
1.213e-03 & 4.246e-02 & 7.651e-02 & 6.846e-01 & 4.827e+00 & 6.793e-03 & 3.046e+02 & 6.839e-08 & 1.922e+04 & 9.800e-13\\
1.335e-03 & 4.577e-02 & 8.425e-02 & 6.993e-01 & 5.315e+00 & 5.419e-03 & 3.354e+02 & 5.236e-08 & 2.116e+04 & 7.597e-13\\
1.470e-03 & 4.934e-02 & 9.277e-02 & 7.104e-01 & 5.853e+00 & 4.306e-03 & 3.693e+02 & 4.012e-08 & 2.330e+04 & 5.891e-13\\
1.619e-03 & 5.318e-02 & 1.022e-01 & 7.175e-01 & 6.446e+00 & 3.408e-03 & 4.067e+02 & 3.076e-08 & 2.566e+04 & 4.568e-13\\
1.783e-03 & 5.733e-02 & 1.125e-01 & 7.206e-01 & 7.098e+00 & 2.688e-03 & 4.478e+02 & 2.361e-08 & 2.825e+04 & 3.543e-13\\
1.963e-03 & 6.181e-02 & 1.239e-01 & 7.199e-01 & 7.816e+00 & 2.111e-03 & 4.931e+02 & 1.812e-08 & 3.111e+04 & 2.749e-13\\
2.162e-03 & 6.663e-02 & 1.364e-01 & 7.156e-01 & 8.607e+00 & 1.652e-03 & 5.430e+02 & 1.393e-08 & 3.426e+04 & 2.133e-13\\
2.381e-03 & 7.182e-02 & 1.502e-01 & 7.078e-01 & 9.478e+00 & 1.288e-03 & 5.980e+02 & 1.071e-08 & 3.773e+04 & 1.656e-13\\
2.622e-03 & 7.741e-02 & 1.654e-01 & 6.968e-01 & 1.044e+01 & 1.001e-03 & 6.585e+02 & 8.237e-09 & 4.155e+04 & 1.285e-13\\
2.887e-03 & 8.343e-02 & 1.822e-01 & 6.826e-01 & 1.149e+01 & 7.750e-04 & 7.251e+02 & 6.338e-09 & 4.575e+04 & 9.977e-14\\
3.179e-03 & 8.990e-02 & 2.006e-01 & 6.654e-01 & 1.266e+01 & 5.982e-04 & 7.985e+02 & 4.878e-09 & 5.038e+04 & 7.747e-14\\
3.501e-03 & 9.686e-02 & 2.209e-01 & 6.450e-01 & 1.394e+01 & 4.606e-04 & 8.793e+02 & 3.755e-09 & 5.548e+04 & 6.017e-14\\
3.855e-03 & 1.043e-01 & 2.432e-01 & 6.218e-01 & 1.535e+01 & 3.538e-04 & 9.682e+02 & 2.892e-09 & 6.109e+04 & 4.674e-14\\
4.245e-03 & 1.124e-01 & 2.678e-01 & 5.957e-01 & 1.690e+01 & 2.712e-04 & 1.066e+03 & 2.228e-09 & 6.727e+04 & 3.631e-14\\
4.675e-03 & 1.210e-01 & 2.949e-01 & 5.671e-01 & 1.861e+01 & 2.075e-04 & 1.174e+03 & 1.717e-09 & 7.408e+04 & 2.821e-14\\
5.148e-03 & 1.303e-01 & 3.248e-01 & 5.360e-01 & 2.049e+01 & 1.585e-04 & 1.293e+03 & 1.324e-09 & 8.157e+04 & 2.192e-14\\
5.669e-03 & 1.402e-01 & 3.577e-01 & 5.027e-01 & 2.257e+01 & 1.208e-04 & 1.424e+03 & 1.021e-09 & 8.983e+04 & 1.704e-14\\
6.242e-03 & 1.508e-01 & 3.938e-01 & 4.679e-01 & 2.485e+01 & 9.194e-05 & 1.568e+03 & 7.873e-10 & 9.892e+04 & 1.325e-14\\
6.874e-03 & 1.621e-01 & 4.337e-01 & 4.318e-01 & 2.736e+01 & 6.988e-05 & 1.726e+03 & 6.073e-10 & \nodata & \nodata\\
7.569e-03 & 1.743e-01 & 4.776e-01 & 3.953e-01 & 3.013e+01 & 5.305e-05 & 1.901e+03 & 4.685e-10 & \nodata & \nodata\\
8.335e-03 & 1.872e-01 & 5.259e-01 & 3.592e-01 & 3.318e+01 & 4.023e-05 & 2.093e+03 & 3.615e-10 & \nodata & \nodata\\
9.179e-03 & 2.010e-01 & 5.791e-01 & 3.239e-01 & 3.654e+01 & 3.048e-05 & 2.305e+03 & 2.790e-10 & \nodata & \nodata\\
1.011e-02 & 2.157e-01 & 6.377e-01 & 2.897e-01 & 4.023e+01 & 2.308e-05 & 2.539e+03 & 2.154e-10 & \nodata & \nodata\\
1.113e-02 & 2.313e-01 & 7.022e-01 & 2.570e-01 & 4.431e+01 & 1.746e-05 & 2.795e+03 & 1.663e-10 & \nodata & \nodata\\
1.226e-02 & 2.479e-01 & 7.733e-01 & 2.261e-01 & 4.879e+01 & 1.320e-05 & 3.078e+03 & 1.284e-10 & \nodata & \nodata\\
1.350e-02 & 2.654e-01 & 8.515e-01 & 1.974e-01 & 5.373e+01 & 9.968e-06 & 3.390e+03 & 9.916e-11 & \nodata & \nodata\\
1.486e-02 & 2.840e-01 & 9.377e-01 & 1.712e-01 & 5.916e+01 & 7.525e-06 & 3.733e+03 & 7.660e-11 & \nodata & \nodata\\
1.637e-02 & 3.035e-01 & 1.033e+00 & 1.474e-01 & 6.515e+01 & 5.679e-06 & 4.110e+03 & 5.918e-11 & \nodata & \nodata\\
1.802e-02 & 3.240e-01 & 1.137e+00 & 1.260e-01 & 7.174e+01 & 4.286e-06 & 4.526e+03 & 4.573e-11 & \nodata & \nodata\\
1.985e-02 & 3.456e-01 & 1.252e+00 & 1.070e-01 & 7.900e+01 & 3.235e-06 & 4.984e+03 & 3.535e-11 & \nodata & \nodata\\
2.185e-02 & 3.682e-01 & 1.379e+00 & 9.027e-02 & 8.699e+01 & 2.442e-06 & 5.489e+03 & 2.732e-11 & \nodata & \nodata\\
2.406e-02 & 3.917e-01 & 1.518e+00 & 7.579e-02 & 9.580e+01 & 1.844e-06 & 6.044e+03 & 2.113e-11 & \nodata & \nodata\\
2.650e-02 & 4.161e-01 & 1.672e+00 & 6.338e-02 & 1.055e+02 & 1.394e-06 & 6.656e+03 & 1.634e-11 & \nodata & \nodata\\
2.918e-02 & 4.414e-01 & 1.841e+00 & 5.278e-02 & 1.162e+02 & 1.054e-06 & 7.329e+03 & 1.264e-11 & \nodata & \nodata\\
3.213e-02 & 4.673e-01 & 2.027e+00 & 4.378e-02 & 1.279e+02 & 7.972e-07 & 8.071e+03 & 9.778e-12 & \nodata & \nodata\\
3.539e-02 & 4.938e-01 & 2.233e+00 & 3.620e-02 & 1.409e+02 & 6.039e-07 & 8.887e+03 & 7.567e-12 & \nodata & \nodata\\
	\enddata
	\tablecomments{Differential Intensity units: (m$^2$ s sr GeV/n)$^{-1}$.}
\end{deluxetable*}

\begin{deluxetable*}{cccccccccc}%[p!]
	\tabletypesize{\footnotesize}
	\tablecolumns{10}
	\tablewidth{0mm}
	\tablecaption{$Z=13$ -- Default aluminum LIS\label{Tbl-BasicAluminumLIS-Rigi}}
	\tablehead{
		\colhead{Rigidity} & \colhead{Differential} &
		\colhead{Rigidity} & \colhead{Differential} &
		\colhead{Rigidity} & \colhead{Differential} &
		\colhead{Rigidity} & \colhead{Differential} &
		\colhead{Rigidity} & \colhead{Differential} \\ [-2ex] 
		\colhead{GV} & \colhead{intensity} &
		\colhead{GV} & \colhead{intensity} &
		\colhead{GV} & \colhead{intensity} &
		\colhead{GV} & \colhead{intensity} &
		\colhead{GV} & \colhead{intensity}  
	}
	\startdata
8.967e-02 & 1.397e-03 & 7.240e-01 & 1.196e-01 & 1.002e+01 & 5.553e-03 & 5.236e+02 & 1.016e-07 & 3.291e+04 & 4.657e-12\\
9.410e-02 & 1.595e-03 & 7.610e-01 & 1.249e-01 & 1.087e+01 & 4.530e-03 & 5.763e+02 & 7.925e-08 & 3.624e+04 & 3.713e-12\\
9.875e-02 & 1.801e-03 & 8.000e-01 & 1.299e-01 & 1.180e+01 & 3.680e-03 & 6.345e+02 & 6.195e-08 & 3.991e+04 & 2.961e-12\\
1.036e-01 & 2.034e-03 & 8.412e-01 & 1.348e-01 & 1.283e+01 & 2.975e-03 & 6.985e+02 & 4.852e-08 & 4.395e+04 & 2.361e-12\\
1.087e-01 & 2.296e-03 & 8.847e-01 & 1.393e-01 & 1.396e+01 & 2.394e-03 & 7.689e+02 & 3.807e-08 & 4.840e+04 & 1.882e-12\\
1.141e-01 & 2.592e-03 & 9.306e-01 & 1.435e-01 & 1.520e+01 & 1.918e-03 & 8.466e+02 & 2.990e-08 & 5.329e+04 & 1.501e-12\\
1.198e-01 & 2.926e-03 & 9.791e-01 & 1.473e-01 & 1.656e+01 & 1.530e-03 & 9.320e+02 & 2.352e-08 & 5.868e+04 & 1.197e-12\\
1.257e-01 & 3.302e-03 & 1.030e+00 & 1.506e-01 & 1.806e+01 & 1.216e-03 & 1.026e+03 & 1.852e-08 & 6.462e+04 & 9.548e-13\\
1.319e-01 & 3.726e-03 & 1.085e+00 & 1.534e-01 & 1.972e+01 & 9.622e-04 & 1.130e+03 & 1.461e-08 & 7.116e+04 & 7.615e-13\\
1.384e-01 & 4.204e-03 & 1.142e+00 & 1.556e-01 & 2.153e+01 & 7.589e-04 & 1.244e+03 & 1.153e-08 & 7.836e+04 & 6.074e-13\\
1.453e-01 & 4.742e-03 & 1.203e+00 & 1.572e-01 & 2.353e+01 & 5.964e-04 & 1.370e+03 & 9.105e-09 & 8.629e+04 & 4.845e-13\\
1.524e-01 & 5.348e-03 & 1.268e+00 & 1.582e-01 & 2.573e+01 & 4.670e-04 & 1.508e+03 & 7.196e-09 & 9.502e+04 & 3.865e-13\\
1.600e-01 & 6.029e-03 & 1.336e+00 & 1.586e-01 & 2.815e+01 & 3.646e-04 & 1.660e+03 & 5.691e-09 & 1.046e+05 & 3.083e-13\\
1.679e-01 & 6.795e-03 & 1.409e+00 & 1.584e-01 & 3.082e+01 & 2.839e-04 & 1.828e+03 & 4.503e-09 & 1.152e+05 & 2.459e-13\\
1.762e-01 & 7.655e-03 & 1.487e+00 & 1.578e-01 & 3.375e+01 & 2.205e-04 & 2.013e+03 & 3.565e-09 & 1.269e+05 & 1.962e-13\\
1.849e-01 & 8.620e-03 & 1.569e+00 & 1.562e-01 & 3.698e+01 & 1.709e-04 & 2.216e+03 & 2.824e-09 & 1.397e+05 & 1.565e-13\\
1.941e-01 & 9.703e-03 & 1.657e+00 & 1.538e-01 & 4.054e+01 & 1.322e-04 & 2.440e+03 & 2.239e-09 & 1.539e+05 & 1.249e-13\\
2.037e-01 & 1.092e-02 & 1.751e+00 & 1.504e-01 & 4.445e+01 & 1.020e-04 & 2.687e+03 & 1.775e-09 & 1.694e+05 & 9.962e-14\\
2.138e-01 & 1.227e-02 & 1.851e+00 & 1.460e-01 & 4.876e+01 & 7.861e-05 & 2.959e+03 & 1.408e-09 & 1.866e+05 & 7.947e-14\\
2.243e-01 & 1.379e-02 & 1.958e+00 & 1.407e-01 & 5.351e+01 & 6.045e-05 & 3.258e+03 & 1.117e-09 & 2.054e+05 & 6.341e-14\\
2.355e-01 & 1.549e-02 & 2.073e+00 & 1.344e-01 & 5.873e+01 & 4.642e-05 & 3.588e+03 & 8.870e-10 & \nodata & \nodata\\
2.471e-01 & 1.738e-02 & 2.196e+00 & 1.273e-01 & 6.449e+01 & 3.560e-05 & 3.950e+03 & 7.043e-10 & \nodata & \nodata\\
2.594e-01 & 1.948e-02 & 2.328e+00 & 1.195e-01 & 7.082e+01 & 2.727e-05 & 4.350e+03 & 5.594e-10 & \nodata & \nodata\\
2.723e-01 & 2.182e-02 & 2.470e+00 & 1.111e-01 & 7.780e+01 & 2.086e-05 & 4.790e+03 & 4.445e-10 & \nodata & \nodata\\
2.858e-01 & 2.442e-02 & 2.623e+00 & 1.022e-01 & 8.548e+01 & 1.595e-05 & 5.274e+03 & 3.533e-10 & \nodata & \nodata\\
3.000e-01 & 2.729e-02 & 2.788e+00 & 9.325e-02 & 9.394e+01 & 1.218e-05 & 5.808e+03 & 2.808e-10 & \nodata & \nodata\\
3.149e-01 & 3.047e-02 & 2.965e+00 & 8.441e-02 & 1.032e+02 & 9.299e-06 & 6.395e+03 & 2.233e-10 & \nodata & \nodata\\
3.305e-01 & 3.397e-02 & 3.158e+00 & 7.574e-02 & 1.135e+02 & 7.091e-06 & 7.042e+03 & 1.776e-10 & \nodata & \nodata\\
3.470e-01 & 3.782e-02 & 3.366e+00 & 6.736e-02 & 1.248e+02 & 5.404e-06 & 7.754e+03 & 1.413e-10 & \nodata & \nodata\\
3.642e-01 & 4.202e-02 & 3.591e+00 & 5.940e-02 & 1.372e+02 & 4.117e-06 & 8.539e+03 & 1.124e-10 & \nodata & \nodata\\
3.824e-01 & 4.657e-02 & 3.836e+00 & 5.194e-02 & 1.509e+02 & 3.136e-06 & 9.403e+03 & 8.946e-11 & \nodata & \nodata\\
4.015e-01 & 5.143e-02 & 4.102e+00 & 4.505e-02 & 1.660e+02 & 2.389e-06 & 1.035e+04 & 7.121e-11 & \nodata & \nodata\\
4.215e-01 & 5.654e-02 & 4.391e+00 & 3.880e-02 & 1.826e+02 & 1.821e-06 & 1.140e+04 & 5.669e-11 & \nodata & \nodata\\
4.426e-01 & 6.191e-02 & 4.706e+00 & 3.322e-02 & 2.009e+02 & 1.389e-06 & 1.255e+04 & 4.514e-11 & \nodata & \nodata\\
4.647e-01 & 6.747e-02 & 5.049e+00 & 2.830e-02 & 2.210e+02 & 1.060e-06 & 1.383e+04 & 3.595e-11 & \nodata & \nodata\\
4.880e-01 & 7.321e-02 & 5.424e+00 & 2.400e-02 & 2.432e+02 & 8.097e-07 & 1.522e+04 & 2.863e-11 & \nodata & \nodata\\
5.125e-01 & 7.907e-02 & 5.833e+00 & 2.027e-02 & 2.676e+02 & 6.196e-07 & 1.676e+04 & 2.281e-11 & \nodata & \nodata\\
5.383e-01 & 8.500e-02 & 6.280e+00 & 1.707e-02 & 2.945e+02 & 4.750e-07 & 1.846e+04 & 1.817e-11 & \nodata & \nodata\\
	\enddata
	\tablecomments{Differential Intensity units: (m$^2$ s sr GV)$^{-1}$.}
\end{deluxetable*}

\begin{deluxetable*}{cccccccccc}%[p!]
	\tabletypesize{\footnotesize}
	\tablecolumns{10}
	\tablewidth{0mm}
	\tablecaption{$Z=13$ -- Default aluminum LIS\label{Tbl-BasicAluminumLIS-EKin}}
	\tablehead{
		\colhead{$E_{\rm kin}$} & \colhead{Differential} &
		\colhead{$E_{\rm kin}$} & \colhead{Differential} &
		\colhead{$E_{\rm kin}$} & \colhead{Differential} &
		\colhead{$E_{\rm kin}$} & \colhead{Differential} &
		\colhead{$E_{\rm kin}$} & \colhead{Differential} \\ [-2ex] 
		\colhead{GeV/n} & \colhead{intensity} &
		\colhead{GeV/n} & \colhead{intensity} &
		\colhead{GeV/n} & \colhead{intensity} &
		\colhead{GeV/n} & \colhead{intensity} &
		\colhead{GeV/n} & \colhead{intensity}  
	}
	\startdata
1.000e-03 & 6.256e-02 & 6.309e-02 & 7.060e-01 & 3.981e+00 & 1.179e-02 & 2.512e+02 & 2.122e-07 & 1.585e+04 & 9.680e-12\\
1.101e-03 & 6.811e-02 & 6.948e-02 & 7.058e-01 & 4.384e+00 & 9.592e-03 & 2.766e+02 & 1.655e-07 & 1.745e+04 & 7.717e-12\\
1.213e-03 & 7.329e-02 & 7.651e-02 & 7.037e-01 & 4.827e+00 & 7.776e-03 & 3.046e+02 & 1.294e-07 & 1.922e+04 & 6.153e-12\\
1.335e-03 & 7.886e-02 & 8.425e-02 & 6.996e-01 & 5.315e+00 & 6.275e-03 & 3.354e+02 & 1.013e-07 & 2.116e+04 & 4.906e-12\\
1.470e-03 & 8.485e-02 & 9.277e-02 & 6.935e-01 & 5.853e+00 & 5.042e-03 & 3.693e+02 & 7.946e-08 & 2.330e+04 & 3.912e-12\\
1.619e-03 & 9.128e-02 & 1.022e-01 & 6.855e-01 & 6.446e+00 & 4.034e-03 & 4.067e+02 & 6.241e-08 & 2.566e+04 & 3.120e-12\\
1.783e-03 & 9.820e-02 & 1.125e-01 & 6.756e-01 & 7.098e+00 & 3.215e-03 & 4.478e+02 & 4.908e-08 & 2.825e+04 & 2.488e-12\\
1.963e-03 & 1.056e-01 & 1.239e-01 & 6.636e-01 & 7.816e+00 & 2.552e-03 & 4.931e+02 & 3.864e-08 & 3.111e+04 & 1.984e-12\\
2.162e-03 & 1.136e-01 & 1.364e-01 & 6.497e-01 & 8.607e+00 & 2.019e-03 & 5.430e+02 & 3.047e-08 & 3.426e+04 & 1.582e-12\\
2.381e-03 & 1.222e-01 & 1.502e-01 & 6.340e-01 & 9.478e+00 & 1.591e-03 & 5.980e+02 & 2.404e-08 & 3.773e+04 & 1.262e-12\\
2.622e-03 & 1.313e-01 & 1.654e-01 & 6.168e-01 & 1.044e+01 & 1.250e-03 & 6.585e+02 & 1.899e-08 & 4.155e+04 & 1.007e-12\\
2.887e-03 & 1.412e-01 & 1.822e-01 & 5.982e-01 & 1.149e+01 & 9.785e-04 & 7.251e+02 & 1.500e-08 & 4.575e+04 & 8.030e-13\\
3.179e-03 & 1.517e-01 & 2.006e-01 & 5.785e-01 & 1.266e+01 & 7.638e-04 & 7.985e+02 & 1.186e-08 & 5.038e+04 & 6.405e-13\\
3.501e-03 & 1.629e-01 & 2.209e-01 & 5.578e-01 & 1.394e+01 & 5.946e-04 & 8.793e+02 & 9.385e-09 & 5.548e+04 & 5.110e-13\\
3.855e-03 & 1.750e-01 & 2.432e-01 & 5.366e-01 & 1.535e+01 & 4.618e-04 & 9.682e+02 & 7.429e-09 & 6.109e+04 & 4.076e-13\\
4.245e-03 & 1.878e-01 & 2.678e-01 & 5.141e-01 & 1.690e+01 & 3.579e-04 & 1.066e+03 & 5.884e-09 & 6.727e+04 & 3.252e-13\\
4.675e-03 & 2.015e-01 & 2.949e-01 & 4.901e-01 & 1.861e+01 & 2.768e-04 & 1.174e+03 & 4.664e-09 & 7.408e+04 & 2.594e-13\\
5.148e-03 & 2.161e-01 & 3.248e-01 & 4.647e-01 & 2.049e+01 & 2.137e-04 & 1.293e+03 & 3.697e-09 & 8.157e+04 & 2.070e-13\\
5.669e-03 & 2.316e-01 & 3.577e-01 & 4.380e-01 & 2.257e+01 & 1.646e-04 & 1.424e+03 & 2.933e-09 & 8.983e+04 & 1.651e-13\\
6.242e-03 & 2.481e-01 & 3.938e-01 & 4.102e-01 & 2.485e+01 & 1.266e-04 & 1.568e+03 & 2.327e-09 & 9.892e+04 & 1.317e-13\\
6.874e-03 & 2.656e-01 & 4.337e-01 & 3.816e-01 & 2.736e+01 & 9.719e-05 & 1.726e+03 & 1.847e-09 & \nodata & \nodata\\
7.569e-03 & 2.841e-01 & 4.776e-01 & 3.522e-01 & 3.013e+01 & 7.454e-05 & 1.901e+03 & 1.466e-09 & \nodata & \nodata\\
8.335e-03 & 3.036e-01 & 5.259e-01 & 3.226e-01 & 3.318e+01 & 5.709e-05 & 2.093e+03 & 1.165e-09 & \nodata & \nodata\\
9.179e-03 & 3.243e-01 & 5.791e-01 & 2.931e-01 & 3.654e+01 & 4.369e-05 & 2.305e+03 & 9.252e-10 & \nodata & \nodata\\
1.011e-02 & 3.460e-01 & 6.377e-01 & 2.640e-01 & 4.023e+01 & 3.340e-05 & 2.539e+03 & 7.352e-10 & \nodata & \nodata\\
1.113e-02 & 3.688e-01 & 7.022e-01 & 2.359e-01 & 4.431e+01 & 2.551e-05 & 2.795e+03 & 5.844e-10 & \nodata & \nodata\\
1.226e-02 & 3.926e-01 & 7.733e-01 & 2.096e-01 & 4.879e+01 & 1.947e-05 & 3.078e+03 & 4.646e-10 & \nodata & \nodata\\
1.350e-02 & 4.175e-01 & 8.515e-01 & 1.848e-01 & 5.373e+01 & 1.485e-05 & 3.390e+03 & 3.695e-10 & \nodata & \nodata\\
1.486e-02 & 4.433e-01 & 9.377e-01 & 1.617e-01 & 5.916e+01 & 1.131e-05 & 3.733e+03 & 2.939e-10 & \nodata & \nodata\\
1.637e-02 & 4.698e-01 & 1.033e+00 & 1.405e-01 & 6.515e+01 & 8.618e-06 & 4.110e+03 & 2.338e-10 & \nodata & \nodata\\
1.802e-02 & 4.968e-01 & 1.137e+00 & 1.211e-01 & 7.174e+01 & 6.564e-06 & 4.526e+03 & 1.861e-10 & \nodata & \nodata\\
1.985e-02 & 5.235e-01 & 1.252e+00 & 1.037e-01 & 7.900e+01 & 5.001e-06 & 4.984e+03 & 1.481e-10 & \nodata & \nodata\\
2.185e-02 & 5.492e-01 & 1.379e+00 & 8.831e-02 & 8.699e+01 & 3.811e-06 & 5.489e+03 & 1.179e-10 & \nodata & \nodata\\
2.406e-02 & 5.738e-01 & 1.518e+00 & 7.479e-02 & 9.580e+01 & 2.906e-06 & 6.044e+03 & 9.387e-11 & \nodata & \nodata\\
2.650e-02 & 5.969e-01 & 1.672e+00 & 6.311e-02 & 1.055e+02 & 2.217e-06 & 6.656e+03 & 7.475e-11 & \nodata & \nodata\\
2.918e-02 & 6.182e-01 & 1.841e+00 & 5.306e-02 & 1.162e+02 & 1.694e-06 & 7.329e+03 & 5.953e-11 & \nodata & \nodata\\
3.213e-02 & 6.376e-01 & 2.027e+00 & 4.446e-02 & 1.279e+02 & 1.296e-06 & 8.071e+03 & 4.742e-11 & \nodata & \nodata\\
3.539e-02 & 6.548e-01 & 2.233e+00 & 3.721e-02 & 1.409e+02 & 9.936e-07 & 8.887e+03 & 3.778e-11 & \nodata & \nodata\\
	\enddata
	\tablecomments{Differential Intensity units: (m$^2$ s sr GeV/n)$^{-1}$.}
\end{deluxetable*}

\begin{deluxetable*}{cccccccccc}%[p!]
	\tabletypesize{\footnotesize}
	\tablecolumns{10}
	\tablewidth{0mm}
	\tablecaption{$Z=13$ -- Total aluminum LIS\label{Tbl-CompleteAluminumLIS-Rigi}}
	\tablehead{
		\colhead{Rigidity} & \colhead{Differential} &
		\colhead{Rigidity} & \colhead{Differential} &
		\colhead{Rigidity} & \colhead{Differential} &
		\colhead{Rigidity} & \colhead{Differential} &
		\colhead{Rigidity} & \colhead{Differential} \\ [-2ex] 
		\colhead{GV} & \colhead{intensity} &
		\colhead{GV} & \colhead{intensity} &
		\colhead{GV} & \colhead{intensity} &
		\colhead{GV} & \colhead{intensity} &
		\colhead{GV} & \colhead{intensity}  
	}
	\startdata
8.967e-02 & 1.397e-03 & 7.240e-01 & 1.197e-01 & 1.002e+01 & 5.812e-03 & 5.236e+02 & 1.016e-07 & 3.291e+04 & 4.657e-12\\
9.410e-02 & 1.596e-03 & 7.610e-01 & 1.250e-01 & 1.087e+01 & 4.701e-03 & 5.763e+02 & 7.925e-08 & 3.624e+04 & 3.713e-12\\
9.875e-02 & 1.802e-03 & 8.000e-01 & 1.301e-01 & 1.180e+01 & 3.791e-03 & 6.345e+02 & 6.195e-08 & 3.991e+04 & 2.961e-12\\
1.036e-01 & 2.034e-03 & 8.412e-01 & 1.350e-01 & 1.283e+01 & 3.047e-03 & 6.985e+02 & 4.852e-08 & 4.395e+04 & 2.361e-12\\
1.087e-01 & 2.297e-03 & 8.847e-01 & 1.395e-01 & 1.396e+01 & 2.440e-03 & 7.689e+02 & 3.807e-08 & 4.840e+04 & 1.882e-12\\
1.141e-01 & 2.593e-03 & 9.306e-01 & 1.438e-01 & 1.520e+01 & 1.947e-03 & 8.466e+02 & 2.990e-08 & 5.329e+04 & 1.501e-12\\
1.198e-01 & 2.927e-03 & 9.791e-01 & 1.476e-01 & 1.656e+01 & 1.549e-03 & 9.320e+02 & 2.352e-08 & 5.868e+04 & 1.197e-12\\
1.257e-01 & 3.303e-03 & 1.030e+00 & 1.509e-01 & 1.806e+01 & 1.227e-03 & 1.026e+03 & 1.852e-08 & 6.462e+04 & 9.548e-13\\
1.319e-01 & 3.727e-03 & 1.085e+00 & 1.537e-01 & 1.972e+01 & 9.696e-04 & 1.130e+03 & 1.461e-08 & 7.116e+04 & 7.615e-13\\
1.384e-01 & 4.206e-03 & 1.142e+00 & 1.560e-01 & 2.153e+01 & 7.635e-04 & 1.244e+03 & 1.153e-08 & 7.836e+04 & 6.074e-13\\
1.453e-01 & 4.744e-03 & 1.203e+00 & 1.577e-01 & 2.353e+01 & 5.993e-04 & 1.370e+03 & 9.105e-09 & 8.629e+04 & 4.845e-13\\
1.524e-01 & 5.350e-03 & 1.268e+00 & 1.588e-01 & 2.573e+01 & 4.688e-04 & 1.508e+03 & 7.196e-09 & 9.502e+04 & 3.865e-13\\
1.600e-01 & 6.031e-03 & 1.336e+00 & 1.593e-01 & 2.815e+01 & 3.657e-04 & 1.660e+03 & 5.691e-09 & 1.046e+05 & 3.083e-13\\
1.679e-01 & 6.797e-03 & 1.409e+00 & 1.592e-01 & 3.082e+01 & 2.846e-04 & 1.828e+03 & 4.503e-09 & 1.152e+05 & 2.459e-13\\
1.762e-01 & 7.657e-03 & 1.487e+00 & 1.586e-01 & 3.375e+01 & 2.210e-04 & 2.013e+03 & 3.565e-09 & 1.269e+05 & 1.962e-13\\
1.849e-01 & 8.623e-03 & 1.569e+00 & 1.572e-01 & 3.698e+01 & 1.712e-04 & 2.216e+03 & 2.824e-09 & 1.397e+05 & 1.565e-13\\
1.941e-01 & 9.706e-03 & 1.657e+00 & 1.549e-01 & 4.054e+01 & 1.324e-04 & 2.440e+03 & 2.239e-09 & 1.539e+05 & 1.249e-13\\
2.037e-01 & 1.092e-02 & 1.751e+00 & 1.517e-01 & 4.445e+01 & 1.021e-04 & 2.687e+03 & 1.775e-09 & 1.694e+05 & 9.962e-14\\
2.138e-01 & 1.228e-02 & 1.851e+00 & 1.476e-01 & 4.876e+01 & 7.867e-05 & 2.959e+03 & 1.408e-09 & 1.866e+05 & 7.947e-14\\
2.243e-01 & 1.380e-02 & 1.958e+00 & 1.426e-01 & 5.351e+01 & 6.049e-05 & 3.258e+03 & 1.117e-09 & 2.054e+05 & 6.341e-14\\
2.355e-01 & 1.549e-02 & 2.073e+00 & 1.367e-01 & 5.873e+01 & 4.644e-05 & 3.588e+03 & 8.870e-10 & \nodata & \nodata\\
2.471e-01 & 1.738e-02 & 2.196e+00 & 1.300e-01 & 6.449e+01 & 3.561e-05 & 3.950e+03 & 7.043e-10 & \nodata & \nodata\\
2.594e-01 & 1.949e-02 & 2.328e+00 & 1.227e-01 & 7.082e+01 & 2.728e-05 & 4.350e+03 & 5.594e-10 & \nodata & \nodata\\
2.723e-01 & 2.183e-02 & 2.470e+00 & 1.150e-01 & 7.780e+01 & 2.087e-05 & 4.790e+03 & 4.445e-10 & \nodata & \nodata\\
2.858e-01 & 2.443e-02 & 2.623e+00 & 1.070e-01 & 8.548e+01 & 1.595e-05 & 5.274e+03 & 3.533e-10 & \nodata & \nodata\\
3.000e-01 & 2.730e-02 & 2.788e+00 & 9.897e-02 & 9.394e+01 & 1.219e-05 & 5.808e+03 & 2.808e-10 & \nodata & \nodata\\
3.149e-01 & 3.048e-02 & 2.965e+00 & 9.132e-02 & 1.032e+02 & 9.300e-06 & 6.395e+03 & 2.233e-10 & \nodata & \nodata\\
3.305e-01 & 3.398e-02 & 3.158e+00 & 8.404e-02 & 1.135e+02 & 7.092e-06 & 7.042e+03 & 1.776e-10 & \nodata & \nodata\\
3.470e-01 & 3.783e-02 & 3.366e+00 & 7.722e-02 & 1.248e+02 & 5.404e-06 & 7.754e+03 & 1.413e-10 & \nodata & \nodata\\
3.642e-01 & 4.204e-02 & 3.591e+00 & 7.083e-02 & 1.372e+02 & 4.117e-06 & 8.539e+03 & 1.124e-10 & \nodata & \nodata\\
3.824e-01 & 4.659e-02 & 3.836e+00 & 6.448e-02 & 1.509e+02 & 3.136e-06 & 9.403e+03 & 8.946e-11 & \nodata & \nodata\\
4.015e-01 & 5.146e-02 & 4.102e+00 & 5.738e-02 & 1.660e+02 & 2.389e-06 & 1.035e+04 & 7.121e-11 & \nodata & \nodata\\
4.215e-01 & 5.657e-02 & 4.391e+00 & 4.941e-02 & 1.826e+02 & 1.821e-06 & 1.140e+04 & 5.669e-11 & \nodata & \nodata\\
4.426e-01 & 6.194e-02 & 4.706e+00 & 4.167e-02 & 2.009e+02 & 1.389e-06 & 1.255e+04 & 4.514e-11 & \nodata & \nodata\\
4.647e-01 & 6.750e-02 & 5.049e+00 & 3.476e-02 & 2.210e+02 & 1.060e-06 & 1.383e+04 & 3.595e-11 & \nodata & \nodata\\
4.880e-01 & 7.325e-02 & 5.424e+00 & 2.881e-02 & 2.432e+02 & 8.097e-07 & 1.522e+04 & 2.863e-11 & \nodata & \nodata\\
5.125e-01 & 7.911e-02 & 5.833e+00 & 2.378e-02 & 2.676e+02 & 6.196e-07 & 1.676e+04 & 2.281e-11 & \nodata & \nodata\\
5.383e-01 & 8.505e-02 & 6.280e+00 & 1.959e-02 & 2.945e+02 & 4.750e-07 & 1.846e+04 & 1.817e-11 & \nodata & \nodata\\
	\enddata
	\tablecomments{Differential Intensity units: (m$^2$ s sr GV)$^{-1}$.}
\end{deluxetable*}

\begin{deluxetable*}{cccccccccc}%[p!]
	\tabletypesize{\footnotesize}
	\tablecolumns{10}
	\tablewidth{0mm}
	\tablecaption{$Z=13$ -- Total aluminum LIS\label{Tbl-CompleteAluminumLIS-EKin}}
	\tablehead{
		\colhead{$E_{\rm kin}$} & \colhead{Differential} &
		\colhead{$E_{\rm kin}$} & \colhead{Differential} &
		\colhead{$E_{\rm kin}$} & \colhead{Differential} &
		\colhead{$E_{\rm kin}$} & \colhead{Differential} &
		\colhead{$E_{\rm kin}$} & \colhead{Differential} \\ [-2ex] 
		\colhead{GeV/n} & \colhead{intensity} &
		\colhead{GeV/n} & \colhead{intensity} &
		\colhead{GeV/n} & \colhead{intensity} &
		\colhead{GeV/n} & \colhead{intensity} &
		\colhead{GeV/n} & \colhead{intensity}  
	}
	\startdata
1.000e-03 & 6.258e-02 & 6.309e-02 & 7.068e-01 & 3.981e+00 & 1.234e-02 & 2.512e+02 & 2.122e-07 & 1.585e+04 & 9.680e-12\\
1.101e-03 & 6.813e-02 & 6.948e-02 & 7.066e-01 & 4.384e+00 & 9.952e-03 & 2.766e+02 & 1.655e-07 & 1.745e+04 & 7.717e-12\\
1.213e-03 & 7.331e-02 & 7.651e-02 & 7.045e-01 & 4.827e+00 & 8.010e-03 & 3.046e+02 & 1.294e-07 & 1.922e+04 & 6.153e-12\\
1.335e-03 & 7.888e-02 & 8.425e-02 & 7.005e-01 & 5.315e+00 & 6.426e-03 & 3.354e+02 & 1.013e-07 & 2.116e+04 & 4.906e-12\\
1.470e-03 & 8.487e-02 & 9.277e-02 & 6.946e-01 & 5.853e+00 & 5.138e-03 & 3.693e+02 & 7.946e-08 & 2.330e+04 & 3.912e-12\\
1.619e-03 & 9.131e-02 & 1.022e-01 & 6.867e-01 & 6.446e+00 & 4.095e-03 & 4.067e+02 & 6.241e-08 & 2.566e+04 & 3.120e-12\\
1.783e-03 & 9.823e-02 & 1.125e-01 & 6.768e-01 & 7.098e+00 & 3.254e-03 & 4.478e+02 & 4.908e-08 & 2.825e+04 & 2.488e-12\\
1.963e-03 & 1.057e-01 & 1.239e-01 & 6.650e-01 & 7.816e+00 & 2.577e-03 & 4.931e+02 & 3.864e-08 & 3.111e+04 & 1.984e-12\\
2.162e-03 & 1.136e-01 & 1.364e-01 & 6.512e-01 & 8.607e+00 & 2.034e-03 & 5.430e+02 & 3.047e-08 & 3.426e+04 & 1.582e-12\\
2.381e-03 & 1.222e-01 & 1.502e-01 & 6.357e-01 & 9.478e+00 & 1.601e-03 & 5.980e+02 & 2.404e-08 & 3.773e+04 & 1.262e-12\\
2.622e-03 & 1.314e-01 & 1.654e-01 & 6.187e-01 & 1.044e+01 & 1.256e-03 & 6.585e+02 & 1.899e-08 & 4.155e+04 & 1.007e-12\\
2.887e-03 & 1.412e-01 & 1.822e-01 & 6.003e-01 & 1.149e+01 & 9.823e-04 & 7.251e+02 & 1.500e-08 & 4.575e+04 & 8.030e-13\\
3.179e-03 & 1.517e-01 & 2.006e-01 & 5.808e-01 & 1.266e+01 & 7.661e-04 & 7.985e+02 & 1.186e-08 & 5.038e+04 & 6.405e-13\\
3.501e-03 & 1.630e-01 & 2.209e-01 & 5.604e-01 & 1.394e+01 & 5.961e-04 & 8.793e+02 & 9.385e-09 & 5.548e+04 & 5.110e-13\\
3.855e-03 & 1.750e-01 & 2.432e-01 & 5.395e-01 & 1.535e+01 & 4.627e-04 & 9.682e+02 & 7.429e-09 & 6.109e+04 & 4.076e-13\\
4.245e-03 & 1.879e-01 & 2.678e-01 & 5.174e-01 & 1.690e+01 & 3.585e-04 & 1.066e+03 & 5.884e-09 & 6.727e+04 & 3.252e-13\\
4.675e-03 & 2.016e-01 & 2.949e-01 & 4.938e-01 & 1.861e+01 & 2.772e-04 & 1.174e+03 & 4.664e-09 & 7.408e+04 & 2.594e-13\\
5.148e-03 & 2.162e-01 & 3.248e-01 & 4.689e-01 & 2.049e+01 & 2.139e-04 & 1.293e+03 & 3.697e-09 & 8.157e+04 & 2.070e-13\\
5.669e-03 & 2.317e-01 & 3.577e-01 & 4.429e-01 & 2.257e+01 & 1.647e-04 & 1.424e+03 & 2.933e-09 & 8.983e+04 & 1.651e-13\\
6.242e-03 & 2.482e-01 & 3.938e-01 & 4.159e-01 & 2.485e+01 & 1.267e-04 & 1.568e+03 & 2.327e-09 & 9.892e+04 & 1.317e-13\\
6.874e-03 & 2.657e-01 & 4.337e-01 & 3.881e-01 & 2.736e+01 & 9.724e-05 & 1.726e+03 & 1.847e-09 & \nodata & \nodata\\
7.569e-03 & 2.842e-01 & 4.776e-01 & 3.598e-01 & 3.013e+01 & 7.456e-05 & 1.901e+03 & 1.466e-09 & \nodata & \nodata\\
8.335e-03 & 3.038e-01 & 5.259e-01 & 3.314e-01 & 3.318e+01 & 5.711e-05 & 2.093e+03 & 1.165e-09 & \nodata & \nodata\\
9.179e-03 & 3.244e-01 & 5.791e-01 & 3.034e-01 & 3.654e+01 & 4.370e-05 & 2.305e+03 & 9.252e-10 & \nodata & \nodata\\
1.011e-02 & 3.461e-01 & 6.377e-01 & 2.762e-01 & 4.023e+01 & 3.340e-05 & 2.539e+03 & 7.352e-10 & \nodata & \nodata\\
1.113e-02 & 3.689e-01 & 7.022e-01 & 2.504e-01 & 4.431e+01 & 2.551e-05 & 2.795e+03 & 5.844e-10 & \nodata & \nodata\\
1.226e-02 & 3.928e-01 & 7.733e-01 & 2.267e-01 & 4.879e+01 & 1.947e-05 & 3.078e+03 & 4.646e-10 & \nodata & \nodata\\
1.350e-02 & 4.176e-01 & 8.515e-01 & 2.050e-01 & 5.373e+01 & 1.485e-05 & 3.390e+03 & 3.695e-10 & \nodata & \nodata\\
1.486e-02 & 4.435e-01 & 9.377e-01 & 1.853e-01 & 5.916e+01 & 1.131e-05 & 3.733e+03 & 2.939e-10 & \nodata & \nodata\\
1.637e-02 & 4.701e-01 & 1.033e+00 & 1.674e-01 & 6.515e+01 & 8.619e-06 & 4.110e+03 & 2.338e-10 & \nodata & \nodata\\
1.802e-02 & 4.970e-01 & 1.137e+00 & 1.503e-01 & 7.174e+01 & 6.565e-06 & 4.526e+03 & 1.861e-10 & \nodata & \nodata\\
1.985e-02 & 5.238e-01 & 1.252e+00 & 1.321e-01 & 7.900e+01 & 5.001e-06 & 4.984e+03 & 1.481e-10 & \nodata & \nodata\\
2.185e-02 & 5.495e-01 & 1.379e+00 & 1.124e-01 & 8.699e+01 & 3.811e-06 & 5.489e+03 & 1.179e-10 & \nodata & \nodata\\
2.406e-02 & 5.741e-01 & 1.518e+00 & 9.378e-02 & 9.580e+01 & 2.906e-06 & 6.044e+03 & 9.387e-11 & \nodata & \nodata\\
2.650e-02 & 5.972e-01 & 1.672e+00 & 7.750e-02 & 1.055e+02 & 2.217e-06 & 6.656e+03 & 7.475e-11 & \nodata & \nodata\\
2.918e-02 & 6.186e-01 & 1.841e+00 & 6.368e-02 & 1.162e+02 & 1.694e-06 & 7.329e+03 & 5.953e-11 & \nodata & \nodata\\
3.213e-02 & 6.380e-01 & 2.027e+00 & 5.215e-02 & 1.279e+02 & 1.296e-06 & 8.071e+03 & 4.742e-11 & \nodata & \nodata\\
3.539e-02 & 6.553e-01 & 2.233e+00 & 4.269e-02 & 1.409e+02 & 9.936e-07 & 8.887e+03 & 3.778e-11 & \nodata & \nodata\\
	\enddata
	\tablecomments{Differential Intensity units: (m$^2$ s sr GeV/n)$^{-1}$.}
\end{deluxetable*}

\end{document}